\documentclass[11pt,a4paper]{article}
\usepackage{jcappub}
\usepackage{bm}
\usepackage{subcaption}
\usepackage{tcolorbox}

\usepackage{ctable}

\newcommand{\kNL}{k_{NL}}

\newcommand{\DiracD}{\delta_{\text{D}}}

\newcommand{\llangle}{\langle\kern-2\nulldelimiterspace\langle}
\newcommand{\rrangle}{\rangle\kern-2\nulldelimiterspace \rangle}

\newcommand{\vect}[1]{\bm{\mathrm{{#1}}}}

\newcommand{\Mpc}{\text{Mpc}}

\newcommand{\DA}{D_A}
\newcommand{\DB}{D_B}
\newcommand{\DD}{D_D}
\newcommand{\DE}{D_E}
\newcommand{\DF}{D_F}
\newcommand{\DG}{D_G}
\renewcommand{\DJ}{D_J}

\newcommand{\LegendreP}{\mathcal{P}}

\begin{document}

\title{The unequal‐time matter power spectrum: impact on weak lensing observables}
\author[a]{Lucia F. de la Bella,} \author[a,b]{Nicolas Tessore} \author[a]{and Sarah Bridle}
\affiliation[a]{Department of Physics and Astronomy,\\
University of Manchester, Oxford Road, Manchester M13 9PL, UK.}
\affiliation[b]{Department of Physics and Astronomy,\\
University College London, Gower Street, London WC1E 6BT, UK.}

\emailAdd{lucia.fonsecadelabella@manchester.ac.uk}

\abstract{ 
We investigate the impact of a common approximation of weak lensing power spectra: the use of single-epoch matter power spectra in integrals over redshift. 
We disentangle this from the closely connected Limber's approximation.
We derive the unequal-time matter power spectrum at one-loop in standard perturbation theory and effective field theory to deal with non-linear physics. We compare these formalisms and conclude that the unequal-time power spectrum using effective field theory breaks for larger scales. As an alternative we introduce the midpoint approximation.
We also provide, for the first time, a fitting function for the time evolution of the effective field theory counterterms based on the Quijote simulations.
Then we compute the angular power spectrum using a 
range of approaches: the Limber approximation, and the geometric and midpoint approximations. We compare our results with the exact calculation at all angular scales using the unequal-time power spectrum.
We use DES Y1 and LSST-like 
redshift distributions
for our analysis. 
We find that the use of the Limber's approximation in weak lensing diverges from the exact calculation of the angular power spectrum on large-angle separations, $\ell < 10$. Even though this deviation is of order $2\%$ maximum for cosmic lensing, we find the biggest effect for galaxy clustering and galaxy-galaxy lensing. We show that not only is this true for upcoming galaxy surveys, but also for current data such as DES Y1. 
Finally, we make our pipeline and analysis publicly available as a Python package called unequalpy.
}

\maketitle

\section{Introduction}\label{sec1}

Cosmology is living a golden era of high-precision observations with upcoming galaxy surveys probing the Universe on unprecedented smaller scales, such as DESI\footnote{\url{https://www.desi.lbl.gov}} , Euclid\footnote{\url{https://www.euclid-ec.org}}, 
the Rubin Observatory 
\footnote{Formerly known as LSST, \url{https://www.lsst.org}} and the Nancy Grace Roman Space Telescope\footnote{Formerly known as WFIRST, \url{https://roman.gsfc.nasa.gov}}. This upcoming galaxy data will have implications in the prediction of cosmic lensing observables.

\begin{figure}[t!]
    \centering
    \includegraphics[width=0.8\linewidth,origin=c]{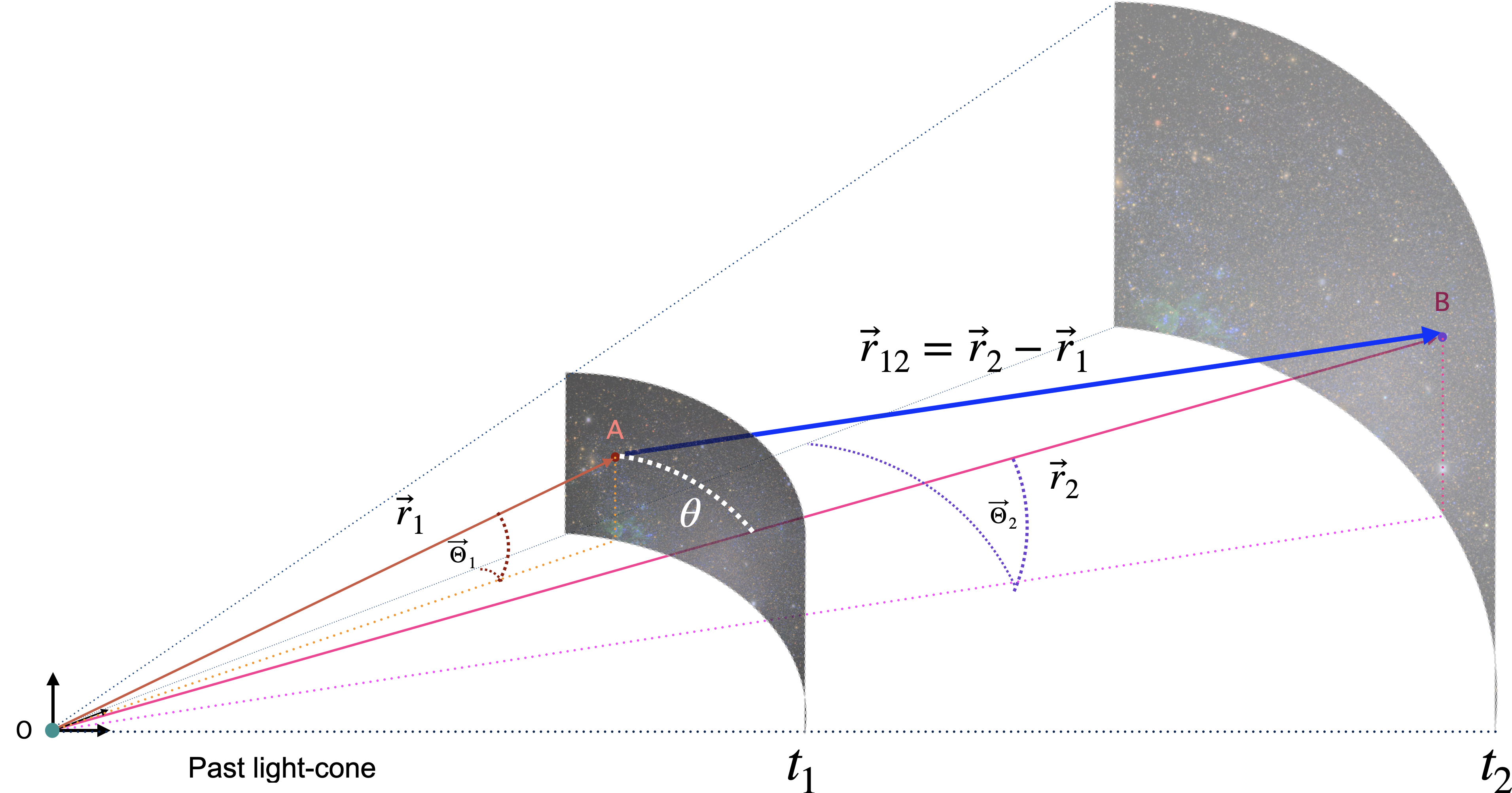}
    \caption{Observations within our past light-cone. This diagram represents two different time slices within our past light-cone, $t_1$ and $t_2$ where $t_1 > t_2$. It shows two source points, $A$ and $B$, at different cosmological distances, $r_1$ and $r_2$ (or redshift, $z_1$ and $z_2$). Note that structure has evolved between these two time slices. 
    }
    \label{fig:Look-back time}
\end{figure}

 When we look at the sky, we observe a two-dimensional projection of different source points at different cosmological distances. In doing so, we look back in time, capturing all the light from our past light cone, see Figure \ref{fig:Look-back time}, observing objects at different time (or redshift) slices\footnote{Note that the observed large-scale structure does not display full spatial symmetry because all observations are done within our past light-cone, breaking homogeneity along the line of sight, and preserving the spherical symmetry on the two-dimensional sky.}. Objects that seem close to each other are actually farther apart and might belong to different time slices. In addition, if the distance between two objects is not large, their look-back time can be thought to be equal. This is the widely-used thin-shell approximation, whose validity should be tested in this era of high-precision cosmology. To do so, we explore the angular correlation functions and the angular power spectrum for different quantities in weak lensing. For two general fields, $A$ and $B$, the angular correlation function reads
\begin{equation}\label{eq: angular_correlation_function_AB}
    \omega^{(i,j)}_{AB}(\theta) = \iint dr_1 dr_2 f_A^i(r_1)f_B^j(r_2) \xi_{AB}(r_{12};r_1,r_2) 
\end{equation}
with $f^k_C(r)$ the weight functions for redshift bins $k$. $\xi_{AB}(r_{12};r_1,r_2)$ is the spatial unequal-time correlation function for spatial separations $r_{12}$, c.f. Fig. \ref{fig:Look-back time}. This quantity measures the correlation between the fields $A$ and $B$ at two different time slices and its Fourier transform is the unequal-time power spectrum, $P_{AB}(k; r_1,r_2)$. 
Likewise, after expanding in spherical harmonics, the angular power spectrum yields
\begin{equation}\label{eq: angular_power_spectrum_AB}
    C^{(i,j)}_{AB}(\ell) =  \frac{2}{\pi} \int  dk k^2 \iint dr_1 dr_2 f^i_A(r_1)f^j_B(r_2) j_{\ell}(kr_1)j_{\ell}(kr_2) P_{AB}(k;r_1,r_2) \; .
\end{equation}
with $j_{\ell}(kr)$ the spherical Bessel function of order $\ell$. For a detailed derivation, the reader can refer to \cite{Kitching:2016xcl}, for example. Solving these integrals can be non-trivial, mostly when working in harmonic space where the spherical Bessel functions present a highly-oscillatory behaviour. Through history, many numerical and analytical approaches have been tested. For a summary on the numerical approaches, one could refer to our accompanying paper \cite{Tessore_2020b}. On the analytical side, one could focus on approximations to the unequal-time correlators, or make assumptions about the filters or approximate the spherical Bessel functions to the amplitude of their first peak.\\

At the level of the power spectrum, the most widely-used approximation to the unequal-time correlators is the geometric approximation. For this one computes the geometric mean of the two equal-time power spectra  at comoving distances or redshifts $z_1$ and $z_2$, respectively
\begin{equation}\label{eq:geometric_mean}
    P(k; z_1, z_2) \approx \sqrt{P(k; z_1)P(k; z_2)}\; .
\end{equation}
This means that instead of obtaining the full correlation \textit{between} two time slices, Fig. \ref{fig:Look-back time}, one simply computes the power spectrum \textit{at} each time slice . The validity of this approximation is restricted to very large scales where linear physics is exact. That being said, there exist some work on the computation of unequal-time correlators. Kitching $et\, al.$ \cite{Kitching:2016xcl} were among the first ones in deriving such description using one-loop standard perturbation theory. They address the accuracy issue for the equal-time power spectrum approximation, claiming that even the use of the geometric mean power spectrum \eqref{eq:geometric_mean} may result in biased predictions of the cosmic shear power spectrum for Euclid- or LSST-like weak lensing experiments. They stress that in order to compute unequal-time correlators to sufficient accuracy, advance in perturbation theory on non-linear scales is required. This is part of the main focus of this work.
Conversely, Chisari and Pontzen in \cite{Chisari:2019tig} show that the Zel'dovich approximation in Lagrangian perturbation theory provides a much more accurate ($< 10~\%$) analytical description to model unequal-time correlators and validate their results against N-body simulations.

Developing an accurate description of the unequal-time power spectra is one of the main goals of our work. For this we re-derive the unequal-time power spectrum in standard perturbation theory \cite{Kitching:2016xcl}. We deal with non-linear physics using effective field theory and obtain a new approximation to the unequal-time prescription: the midpoint approximation.\\

At the level of the angular power spectra, Limber made the first attempt to compute the angular correlation functions. In \cite{1953ApJ...117..134L},
they develop 
a method to analyse the counts of the extra-galactic ``nebulae'', i.e. galaxies, in terms of a fluctuating density field for the ``nebulae'' in space. For this, they assume that the filters are smooth and the correlators fall off fast enough so that they can compute all these quantities at the mean redshift. Then they apply their methods to the counts obtained by Shane and Wirtanen at the Lick Observatory \cite{1954ApJ...119..655L}. Twenty years later, Peebles \cite{1973ApJ...185..413P} presents a number of theoretical results for the analysis of data distributed on a sphere, applicable when the survey covers only a portion of the sky, employing a discrete version of the Limber's equation. In the late 90s the approximation starts taking its familiar shape when Kaiser \cite{Kaiser_1998} re-derives the Limber's equation within the flat-sky approximation in Fourier space for a homogeneous and isotropic universe with spatial curvature. A similar derivation for the spherical case can be found in Lemos et al. \cite{Lemos:2017arq}. 

More recently, Simon \cite{Simon:2006gm} revisits the Limber's equation.
They distinguish between two different regimes: small-angle and large-angle separations, showing that Limber’s and the so-called thin-layer equations are approximations for these two extremes. They also study Limber's accuracy for a power-law spatial correlation and claims that Limber’s approximation diverges when galaxies are closely distributed.  This implies that Limber’s equation possibly over-estimates the angular correlation to some degree.
For historical reasons, when employing Limber’s equation, the small-angle approximation is automatically used. Simon explains that this is accurate to about $10\%$ for angles smaller than $\theta \lesssim 40^\circ$. This type of analysis has led to the idea that such small-angle approximations could contribute significantly to the tension between the CMB measurements and weak lensing data ---c.f. \cite{Kitching_2017}. However, Lemos et al. \cite{Lemos:2017arq}  conclude that the impact of small-angle approximations on cosmological parameter estimation is negligible for current data.

In addition, some efforts were devoted to extend the Limber's approximation in its Dirac-delta version. These are the so-called post-Limber approximations. One example is Lo Verde et al. \cite{LoVerde:2008re} where they develop a systematic derivation for the Limber approximation to the angular cross-power spectrum of two random fields as a series expansion in $1/(\ell+1/2)$. Nontheless, it is not clear how this would alleviate the divergence of the series expansion at small $\ell$.

Finally, Fang et al. \cite{Fang:2019xat} present a new method based on a generalised FFTLog algorithm for the efficient computation of angular power spectra beyond the Limber approximation. This simplifies the calculation and improves the numerical speed and stability. They implement their method for galaxy clustering and galaxy-galaxy lensing power spectra, and find that the Limber's approximation for galaxy clustering in future analyses like LSST Year 1 and DES Year 6 may cause significant biases in cosmological parameters, indicating that going beyond the Limber approximation is necessary for these analyses.
One of their key points in their method is the separation of the integrals of the angular power spectrum into large scales and small scales. By doing so, they drop the Limber's approximation in the linear contribution and they never have to compute the unequal-time power spectrum for the non-linear term. However, for the latter, they employ the Limber's approximation that transforms the unequal-time correlation into the usual equal-time power spectrum. Even though this splitting seems natural and logical, we wonder whether they might be losing accuracy on the small-scale contribution and whether the non-linearities are well accounted for. For this reason, we understand that efforts should be focused on implementing an all-angle method to compute the exact unequal-time calculation.

In the following, we develop our analysis on the unequal-time matter power spectrum and its impact on weak lensing. In Section \ref{sec: Unequal-time power spectrum}, we derive the unequal-time matter power spectrum in standard perturbation theory and effective field theory, and we present the new midpoint approximation. For those more interested in weak lensing analysis, we recommend to skip Section \ref{sec: Unequal-time power spectrum} and read Section \ref{sec: Impact on weak lensing observables}. In this section \ref{sec: Impact on weak lensing observables}, we compute the main observables in weak lensing using a range of approximations and the all-angle approach presented in our accompanying paper \cite{Tessore_2020b}. We then analyse the relevance of the unequal-time description using DES Y1, and LSST-like redshift distributions. A brief summary and the main conclusions can be found in Section \ref{sec: Conclusion}, followed by the appendices \ref{app: Full-time power spectrum}, \ref{app: Limber_Geometric} and \ref{app: midpoint}. Our final product is the publicly available python package \verb!unequalpy!\footnote{https://github.com/Lucia-Fonseca/unequalpy.git} \cite{lucia_fonseca_de_la_bella_2020_4268314} with functionality to reproduce all the results and analysis presented in this paper.

\section{Unequal-time power spectrum}
\label{sec: Unequal-time power spectrum}
The main goal of this section is to develop an accurate description of the unequal-time power spectra dealing with non-linear physics. To do so, we re-derive the unequal-time matter power spectrum up to third order in standard perturbation theory \cite{Kitching:2016xcl}. The equal-time power spectrum is defined as
\begin{equation}
    \langle \delta(\vec{k},z) \delta(\vec{k'},z) \rangle \equiv (2\pi)^3 \DiracD(\vec{k}+\vec{k}') P(k;z) \; ,
\end{equation}
whereas the unequal-time power spectrum reads
\begin{equation}\label{eq:definition_uetps}
    \langle \delta(\vec{k},z_1) \delta(\vec{k'},z_2) \rangle \equiv (2\pi)^3 \DiracD(\vec{k}+\vec{k}') P(k;z_1,z_2) \; .
\end{equation}
Note that the definition above \eqref{eq:definition_uetps} ensures an isotropic power spectrum, with no privileged direction in the local coordinate system. Therefore, the power spectrum only depends on the length of the wavenumber $k$ and not on its direction. 

In addition, we deal with non-linear physics using effective field theory. Some authors \cite{Carrasco:2012cv, delaBella:2017qjy} showed the capability of this framework to encode small-scale physics and to provide highly accurate predictions on increasingly smaller scales in the context of equal-time power spectra. To show whether such improvement propagates for unequal-time correlators, we also derive the one-loop unequal-time matter power spectrum using effective field theory of large scale structures.  At the end of this section, we introduce a new approximation to the unequal-time prescription, the midpoint approximation, as an alternative to model non linearities.

\subsection{Standard Perturbation Theory}

To compute the matter power spectrum up to third order in perturbation theory, we need to solve perturbatively the equation of motion for the matter density contrast, $\delta = (\rho - \rho_0)/\rho_0$ (with $\rho$ the matter density field and $\rho_0$ the background density) up to third order. Then we compute the two-point correlation functions using the Einstein--de Sitter approximation, EdS \cite{Bernardeau:2001qr}, and assuming the initial density perturbation $\delta^\ast_{\vect{k}}$ to be a Gaussian
random field. For the curious reader, we also derive the full-time dependence in appendix \ref{app: Full-time power spectrum}. However, for the purpose of our analysis, we stick to the EdS approximation. Then there are three contributions labelled $P_{11}$, $P_{22}$ and $P_{13}$,
\begin{subequations}\label{eq: correlators}
    \begin{align}
        \langle \delta_{\vect{k}_1,1}(z_1) \delta_{\vect{k}_2,1}(z_2) \rangle & = (2\pi)^3 \delta(\vect{k}_1 + \vect{k}_2) P_{11}(k, z_1, z_2) \\
        \langle \delta_{\vect{k}_1,2}(z_1) \delta_{\vect{k}_2,2}(z_2) \rangle & = (2\pi)^3 \delta(\vect{k}_1 + \vect{k}_2) P_{22}(k, z_1, z_2) \\
        \langle \delta_{\vect{k}_1,1}(z_1) \delta_{\vect{k}_2,3}(z_2) + \delta_{\vect{k}_1,3}(z_1) \delta_{\vect{k}_2,1}(z_2) \rangle & = (2\pi)^3 \delta(\vect{k}_1 + \vect{k}_2) 2 P_{13}(k, z_1, z_2) ,
    \end{align}
\end{subequations}
where $k$ is the common magnitude of the wavevectors $\vect{k}_1$ and $\vect{k}_2$.
The linear contribution $P_{11}$ is described as the tree-level power spectrum,
and the sum $P_{22} + 2 P_{13}$ is the one-loop contribution
(c.f. \cite{Bernardeau:2001qr, delaBella:2017qjy} for a detailed expression of these terms). Then the one-loop unequal-time matter power spectrum reads
\begin{equation}\label{eq:SPT_UETC_EdS}
    \begin{split}
    P_{\mathrm{\tiny SPT}}(k; z_1,z_2) & =  D(z_1)D(z_2) P_{11}(k)\\ & + D^2(z_1)D^2(z_2) P_{22}(k) + D(z_1)D(z_2)\left[ D^2(z_1) + D^2(z_2) \right] P_{13}(k)
    \end{split}
\end{equation}
where we factored out the time dependence.  $D(z)$ is the normalised linear growth function, which we compute using \verb!SkyPy 0.3! functions \cite{skypy_collaboration_2020_3755531}.

Finally, the one-loop equal-time power spectrum in standard perturbation theory is retrieved by setting $z_1 = z_2 = z$ in the above definitions \eqref{eq: correlators}. This yields
\begin{equation}\label{eq:SPT_ETC_EdS}
    P_{\mathrm{\tiny SPT}}(k,z) =  D^2(z) P_{11}(k) +  D^4(z) P_{22}(k) +  2 D^4(z) P_{13}(k) \; .
\end{equation}

\subsection{Effective Field Theory}

In order to compute the unequal-time matter power spectrum using effective field theory, we follow the same renormalisation programme described in \citep{delaBella:2017qjy}. We split the 13- and 22-type loop integrals into the linear and the non-linear regimes. The linear contribution is calculated by using the standard perturbation results, where we know this is exact. For the non-linear contribution, we Taylor expand the integrands in terms of $k/ \kNL$, dropping order four contributions. Thus all the microscopic physics is encoded in a redshift-dependent free-fitting parameter, the so-called counterterm, $c(z) \equiv c_s(z) / \kNL$. Then, the unequal-time power spectrum in effective field theory reads
\begin{equation}\label{eq:EFT_UETC_EdS}
    P_{\mathrm{\tiny EFT}}(k;z_1,z_2) =  P_{\mathrm{\tiny SPT}}(k;z_1,z_2) - \frac{1}{2} \left( c^2(z_1) + c^2(z_2) \right) D(z_1) D(z_2) k^2 P_{11}(k)
\end{equation}
Finally, the equal-time power spectrum is retrieved when $z_1 = z_2=z$ in equation \eqref{eq:EFT_UETC_EdS}
\begin{equation}\label{eq:EFT_ETC_EdS}
    P_{\mathrm{\tiny EFT}}(k,z) =  P_{\mathrm{\tiny SPT}}(k,z) - c^2(z) D^2(z) k^2 P_{11}(k) \; .
\end{equation}
In order to fit the counterterms, we use data from the Quijote simulations \cite{Villaescusa-Navarro:2019bje}. These simulations are very well documented, offer a great number of cosmologies, provide the matter power spectrum, and, most importantly, offer enough data for different mean redshifts.

\subsubsection{Time evolution of the counterterms}

The Quijote simulations \cite{Villaescusa-Navarro:2019bje} are a set of 43100 full N-body simulations spanning more than 7000 cosmological models, providing full snapshots of the simulations at redshifts 0, 0.5, 1, 2 and 3. In the following, we explain how we use this set of simulations to fit our counterterms and how we obtain their time evolution.\\

\begin{figure}[t!]
    \begin{subfigure}{0.5\textwidth}
        \centering
        \includegraphics[width=\textwidth]{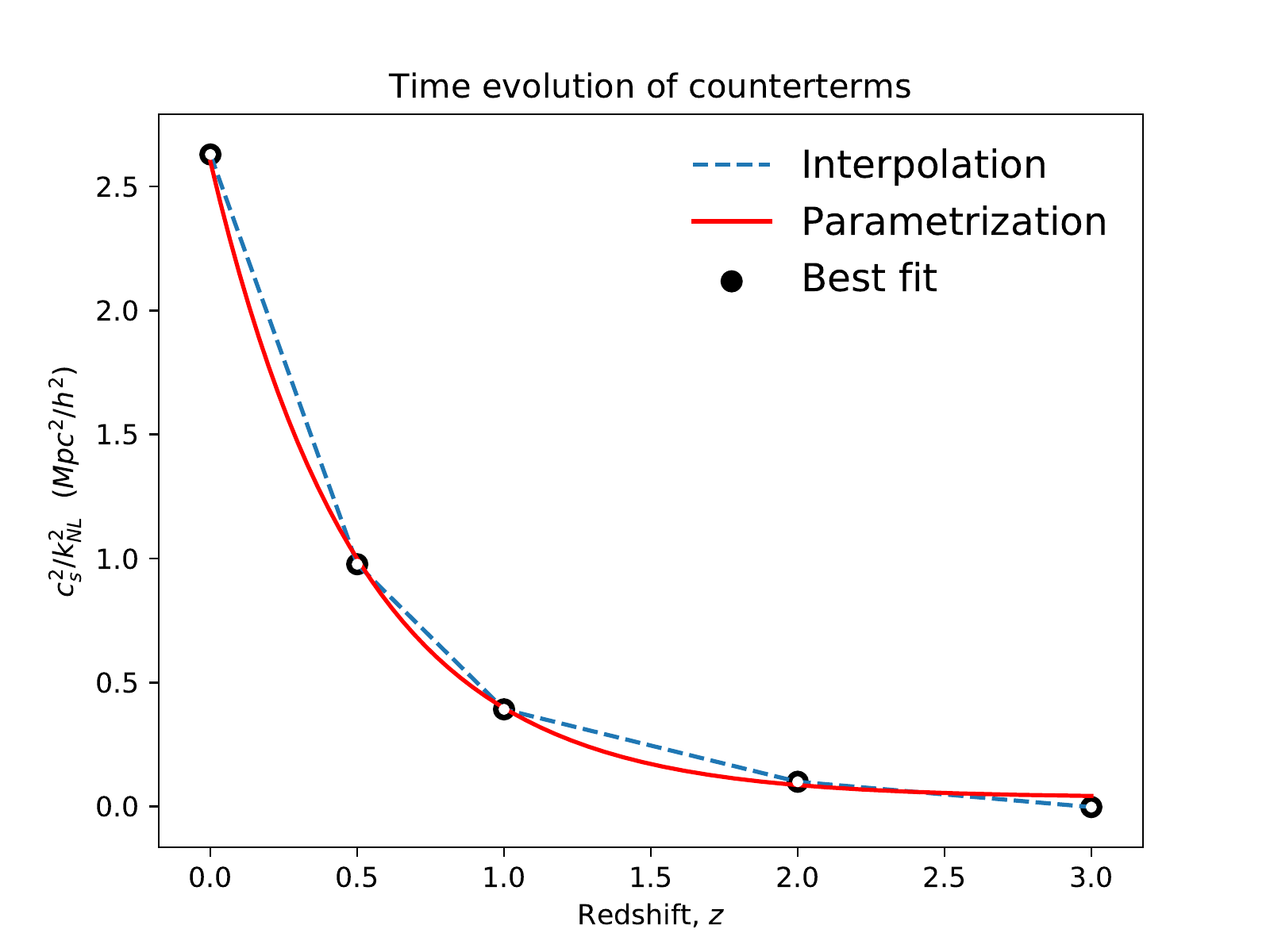}
    \end{subfigure}
    ~~
    \begin{subfigure}{0.5\textwidth}
        \centering
        \includegraphics[width=\textwidth]{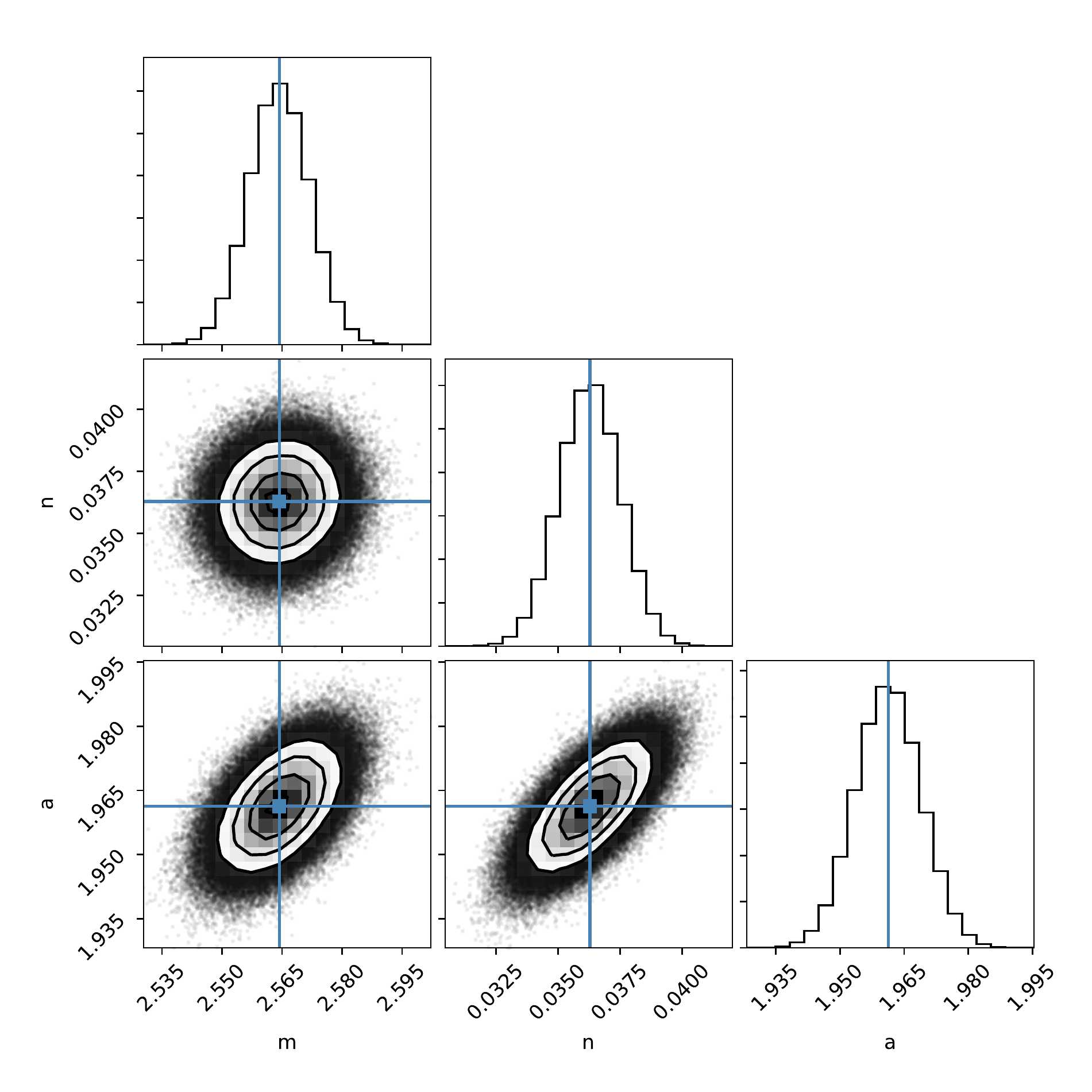}
    \end{subfigure}
    \caption{The left plot shows the parametrization of the counterterms as a function of redshift. The plot on the right represents the correlation between the parameters involved in the time evolution of the counterterms. In this case, the slope and intercept in equation \eqref{eq: counterterms} show the weakest correlation.}
    \label{fig:counterterms}
    \end{figure}

We use the Quijote fiducial model to perform a Bayesian analysis to find the best value of the counterterms \eqref{eq:EFT_ETC_EdS} at every redshift available. We use a flat prior for the free fitting parameter and {\verb!emcee3.0.2!}\footnote{\url{https://emcee.readthedocs.io}}. We use matter power spectrum data up to $k = 0.4 h/\Mpc$, since two-loop contributions start to dominate on smaller scales. The results are shown in Figure \ref{fig:counterterms}. The computed values are well parametrised by a fitting function of the form:
    \begin{equation}
        \label{eq: counterterms}
         c^2(z) = m \textrm{e}^{-a z} + n 
    \end{equation}
    with the best fit values $ m = 2.564^{+0.008}_{- 0.007} \Mpc^2/h^2$, $ n = 0.036^{+0.001}_{- 0.001} \Mpc^2/h^2$ and $a = 1.961^{+0.008}_{- 0.008}$. This is already a new result in cosmology. 
    For example, in \cite{delaBella:2017qjy} authors used a different approach to fit for the counterterms in real space, using an estimator for the fit at every single redshift and using the halofit matter power spectrum given by CAMB\footnote{\url{https://camb.info}}.

In the next section, we analyse which of the above prescriptions provides the best framework to analyse unequal-time correlators, and weak lensing observables.

\subsection{Distinguishing between non-linear and unequal-time effects}
In this section, we assess the performance of the unequal-time matter power spectrum and we explore the effects coming from non-linear physics and unequal-time correlators, c.f. Table \ref{tab: 2x2diagram}. To do so we apply the geometric approximation \eqref{eq:geometric_mean} to the unequal-time matter power spectra. In general,
\begin{equation}\label{eq:geometric_mean_squared}
    P(k;z_1,z_2)^2 = P(k,z_1)P(k,z_2) + \Delta P(k;z_1,z_2)\; .
\end{equation}
where $\Delta P(k;z_1,z_2)$ represents the error when using the geometric approximation. This error vanishes when $z_1 = z_2$ and for linear theory. Equation \eqref{eq:geometric_mean_squared} seems the best description to analyse effects coming from correlations between different time slices. Then we can disentangle non-linear effects by studying the effective field theory counterparts.
\begin{table}[ht]\centering
\begin{tabular}{|c|c|}
\hline
\textbf{\begin{tabular}[c]{@{}c@{}}Equal-time\\ SPT\\ Equation \eqref{eq:SPT_ETC_EdS}\end{tabular}} & \begin{tabular}[c]{@{}c@{}}Equal-time\\ EFT\\Equation \eqref{eq:EFT_ETC_EdS}\end{tabular}           \\ \hline
\begin{tabular}[c]{@{}c@{}}Unequal-time\\ SPT\\Equation \eqref{eq:SPT_UETC_EdS}\end{tabular}         & \textbf{\begin{tabular}[c]{@{}c@{}}Unequal-time\\ EFT\\Equation \eqref{eq:EFT_UETC_EdS}\end{tabular}} \\ \hline
\end{tabular}
\caption{Grid of theoretical descriptions to analyse. SPT stands for ``standard perturbation theory", and EFT for ``effective field theory". Comparing items top to bottom (same column) would describe unequal-time effects, whereas the left-to-right comparison (same row) would describe the effect of the counterterms and non-linear physics. Comparing in diagonal would be a mix of both effects.}
\label{tab: 2x2diagram}
\end{table}

\paragraph{ Non-linear effects:} Left-to-right comparison in Table \ref{tab: 2x2diagram}. We compare standard perturbation theory and effective field theory at the level of the geometric approximation and the full unequal-time correlator. Here we compute the error committed when using standard perturbation theory, instead of effective field theory, both for the geometric approximation and the unequal-time calculation. For both cases the power spectrum of reference is the one from the effective field theory:
    \begin{equation}\label{eq:egeom}
        \epsilon_{geom} = \frac{|P_{\mathrm{\tiny SPT}}(k,z_1)P_{\mathrm{\tiny SPT}}(k,z_2) - P_{\mathrm{\tiny EFT}}(k,z_1)P_{\mathrm{\tiny EFT}}(k,z_2)|}{P_{\mathrm{\tiny EFT}}(k,z_1)P_{\mathrm{\tiny EFT}}(k,z_2)}
    \end{equation}
    \begin{equation}\label{eq:euetc}
        \epsilon_{uet} = \frac{|P_{\mathrm{\tiny SPT}}(k,z_1, z_2)^2 - P_{\mathrm{\tiny EFT}}(k,z_1,z_2)^2|}{P_{\mathrm{\tiny EFT}}(k,z_1,z_2)^2}\;.
    \end{equation}
    
\begin{figure*}[t!]
    \centering
    \includegraphics[width=0.8\textwidth]{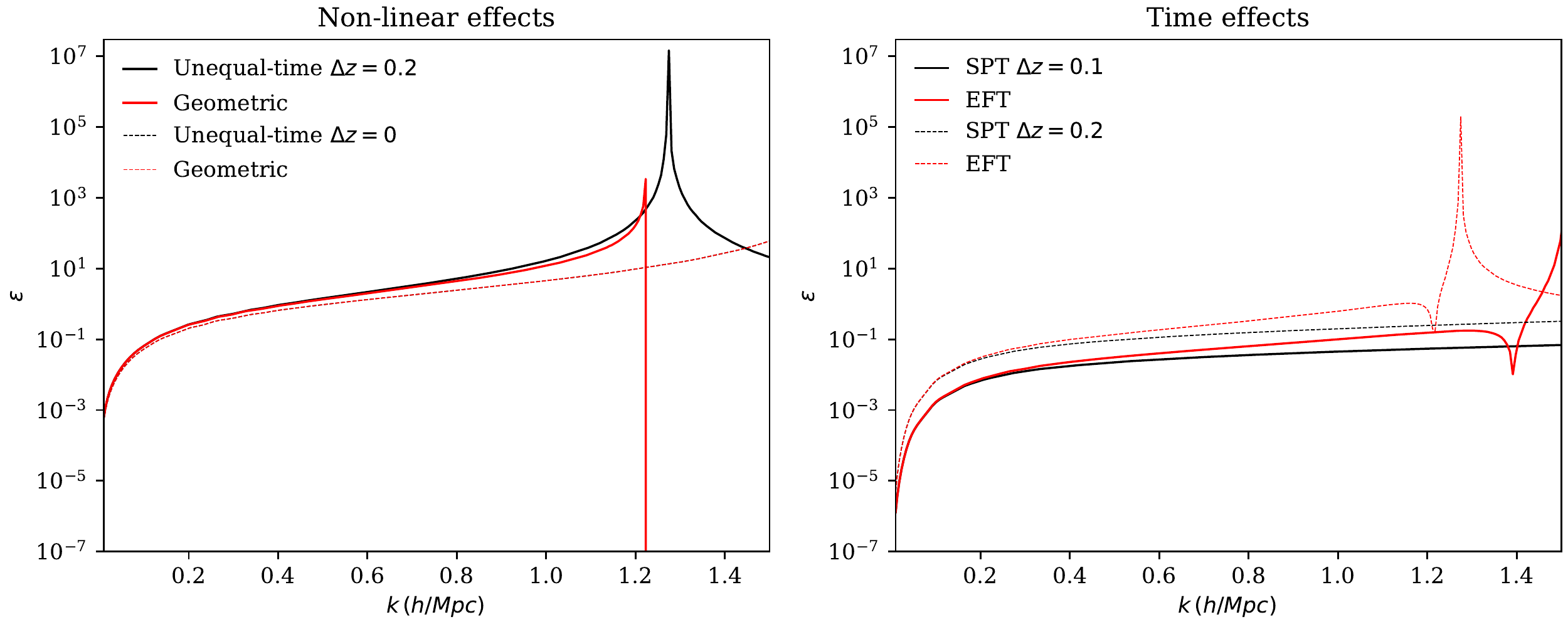}
    \caption{Non-linear effects (left) and unequal-time effects (right) when using the geometric approximation instead of the unequal-time description for both standard perturbation theory and effective field theory.
    On the left, we plot the fractional difference between standard perturbation theory and effective field theory using either the geometric approximation (red lines), equation \eqref{eq:egeom},  or the full unequal-time power spectra (black lines), equation \eqref{eq:euetc}. On the right, we plot the error when using the geometric approximation instead of the unequal-time description (equation \ref{eq:epsilon_theory}) for both standard perturbation theory (black lines) and effective field theory (red lines). This is done for a mean redshift $z_m = 0.5$ and different redshift separations, $\Delta z$.
   The divergences in these plots show where the theory breaks down and higher-order corrections become dominant, therefore the prediction cannot be trusted on those scales. We also observe how this phenomena occurs at lower $k$ values for effective field theory.}
    \label{fig: effects}
\end{figure*}

We calculate this error for a given mean redshift $z_m = 0.5$ and different redshift separations: $\Delta z = 0$ (same time slice) and $\Delta z = 0.2$. In Figure \ref{fig: effects}, we observe that the error increases on the linear regime  and there is little difference between using the geometric and unequal-time calculations. We observe a breaking scale where the prediction diverges which is shifted to lower values of $k$ for higher redshift separations. This breaking scale indicates that higher-order loop corrections are needed (extra counterterms and stochastic parameters), therefore the prediction can no longer be trusted. When we look at the same time slice, $\Delta z = 0$, the absolute error does not vanish, it becomes proportional to the counterterms multiplied by the power spectrum, $\propto 2 c^2(z_m) P_{\mathrm{\tiny SPT}}(k, z_m)$. For increasingly larger redshift separations, the relative error tends to shift to the left and become larger, preserving the shape.

\paragraph{ Unequal-time effects:} Top-to-bottom comparison in Table \ref{tab: 2x2diagram}. We compute the error committed when using the geometric approximation instead of the unequal-time calculation, for both standard perturbation theory and effective field theory. The power spectrum of reference is the unequal-time counterpart
    \begin{equation} \label{eq:epsilon_theory}
        \epsilon_{theory} = \frac{|P_{\mathrm{\tiny theory}}(k,z_1) P_{\mathrm{\tiny theory}}(k,z_2) - P_{\mathrm{\tiny theory}}(k,z_1,z_2)^2|}{P_{\mathrm{\tiny theory}}(k,z_1,z_2)^2}
    \end{equation}
    with $theory = \{SPT, EFT\}$. Again we analyse the error for $z_m=0.5$ and widths $\Delta z = \{0, 0.1,0.2\}$. In Figure \ref{fig: effects}, we observe that the effect of using the geometric approximation instead of using the unequal-time calculation yields an error that increases monotonically. This shows that the geometric approximation is very good on very large scales, where linear theory is exact, but not accurate enough to deal with non-linear physics. In principle, the effective field theory framework should improve such predictions, but it breaks before the SPT counterpart. Again, this effect is due to higher-order effects becoming dominant.  We observe a similar behaviour on large scales for both formalisms, with a larger error for larger redshift separations. For measurements at the same time slice, $\Delta z = 0$, the error vanishes. This is true because the unequal-time correlator equals the geometric approximation when $z_1=z_2$.

In conclusion, Figure \ref{fig: effects} shows the relative comparison between different perturbation formalisms and unequal-time descriptions, c.f. Table \ref{tab: 2x2diagram}. However, answering the question ``which description is best?'' is no trivial task. The effective field theory prediction breaks on intermediate scales and we believe that the current prescription of unequal-time prediction within effective field theory is more sensitive to homogeneity breaking along the line of sight. Therefore we stick with the standard perturbation formalism for the rest of our analysis. 

For the purpose of our weak lensing analysis, we work within the standard framework and develop a new approximation to the unequal-time prescription which alleviates some divergences on the mild non-linear regime. This is the so-called midpoint approximation. 

\subsection{The midpoint approximation}
As a result of our incapability to determine the best prescription of  deal with non-linear physics along the radial direction, we present an alternative and derive the new midpoint approximation, defined as 
\begin{equation}\label{eq:MRA}
    P(k; z_1, z_2) \approx P(k, z_m)
\end{equation}
with $z_m = (z_1 + z_2)/2$. This means that instead of using the continuous information along the line of sight, we only use the power spectrum at the mean redshift, regardless the width $\Delta z$.
There are other choices to define the mean redshift, e.g. the more natural mean on comoving distance, some weighted mean redshift, or choosing the redshift such that you can retrieve the exact growth on large scales. Nonetheless, we make this particular choice for simplicity and the exact definition of the midpoint would have an insignificant effect compared to the use of Limber's approximation or the exact projection method.

Equation \eqref{eq:MRA} is different from the geometric mean approximation \eqref{eq:geometric_mean} which combines the information from the two ends of the redshift shell, $z_1$ and $z_2$. In principle, equation \eqref{eq:MRA} accepts any perturbation theory. We know that in the context of equal-time correlators, effective field theory predictions are several percent levels more accurate than the standard formalism. However, when using effective field theory, the error is larger. Then, until we understand which formalism truly reflects the reality of our universe, we will work within the standard framework. 

\begin{figure*}[ht]
        \centering
        \includegraphics[width=0.8\textwidth]{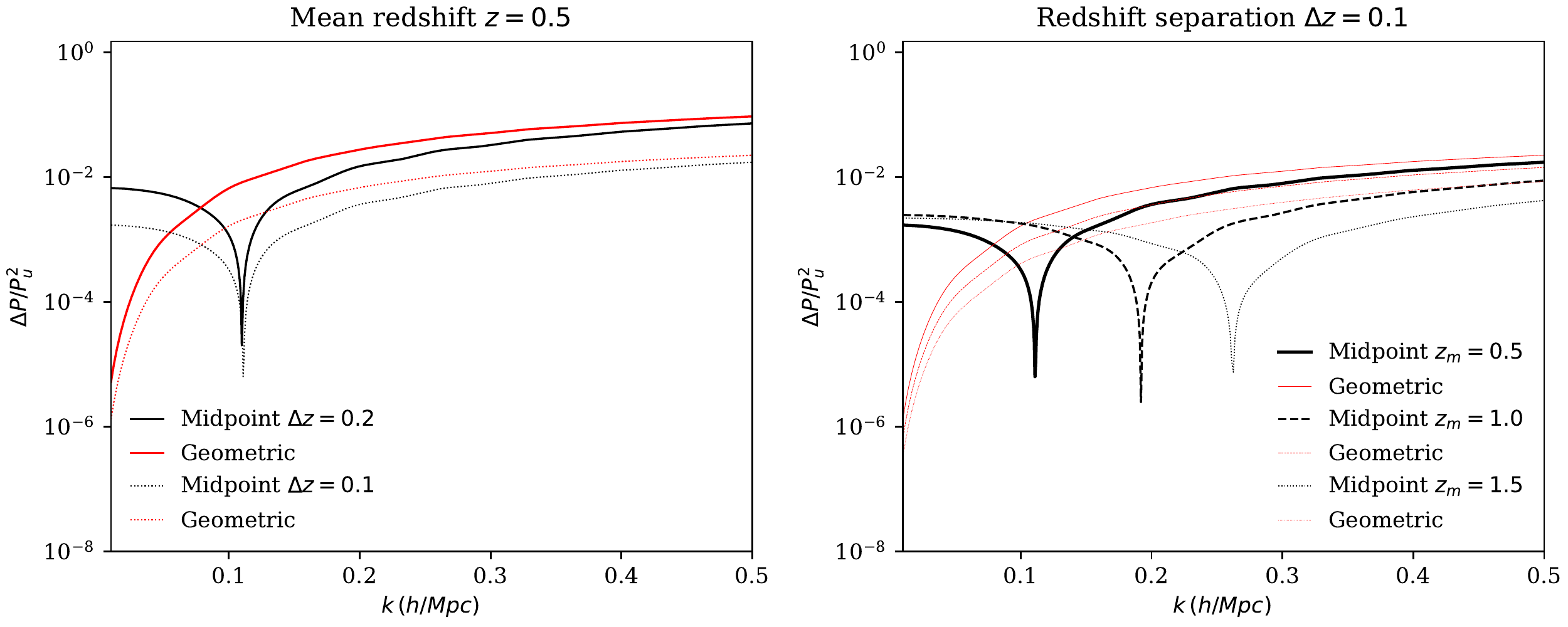}
        \caption{These plots represent the relative error committed when using the traditional geometric approximation (red lines), and the mean redshift approximation in standard perturbation theory (black lines) instead of the unequal-time description, equation. On the left, we plot the error in the squared power spectrum for the midpoint approximation (black lines), equation \eqref{eq:eumid_spt}, and for the geometric approximation (red lines), equation \eqref{eq:epsilon_theory}, for a fixed mean redshift, $z_m=0.5$. It shows the effect of increasing redshift separation between the two epochs and the particular feature for the midpoint approximation at the scale where it equals the prediction from the unequal-time power spectrum. On the right, we show the same error but for a fixed redshift width, $\Delta z=0.1$. It shows the impact of increasing the mean redshift: lower error for larger mean redshift and the midpoint feature shifting to smaller scales.
        } 
        \label{fig: mean approx}
    \end{figure*}

We now show how this new approximation improves our predictions on smaller scales. We compute the error when using one of these approximations instead of the unequal-time power spectrum in standard perturbation theory. For the geometric approximation, this is given by equation \eqref{eq:epsilon_theory}. For the new midpoint approximation, the error reads 
    \begin{equation}\label{eq:eumid_spt}
        \epsilon_{mid}^{spt} = \frac{|P_{\mathrm{\tiny SPT}}(k,z_m)^2 - P_{\mathrm{\tiny SPT}}(k;z_1,z_2)^2|}{P_{\mathrm{\tiny SPT}}(k;z_1,z_2)^2}\;.
    \end{equation}
We show the results in Figure \ref{fig: mean approx}. In the left panel, we fix the midpoint value of redshift $z_m = 0.5$ and consider different redshift separations. Again, we observe how the geometric approximation is exact on very large scales but the error grows on small scales, and again the approximation is worse for larger separations of redshift. The midpoint approximation shows a higher error on larger scales, but the prediction improves on mild non-linear scales. On even smaller scales, the midpoint approximation tends to meet the geometric approximation curve. The new approximation shows a particular feature on a single scale around $k \approx 0.1 h/\Mpc$. This is where the approximation equals the unequal-time value and is characteristic of the midpoint approximation. The geometric approximation only equals the unequal-time value when we look at the same time slice. Beyond such scale, the midpoint approximation underpredicts the actual power spectrum whereas the geometric approximation overpredicts it.

When we fix the redshift separation $\Delta z = 0.1$ and vary the mean value $z_m$ (right panel in Figure \ref{fig: mean approx}), we observe that this turnover happens for increasingly smaller scales as we increase the mean redshift. This means that the unequal-time calculation and the midpoint approximation are equal at increasingly smaller scales for redshift bins that are farther away from the observer. 

In conclusion, the geometric approximation is better on very large scales. We introduced the midpoint approximation hoping that it would predict mild non-linear physics more accurately. However, none of these approximations are completely satisfactory at the level of the matter power spectrum. In the next section, we analyse whether these features propagate when computing weak lensing observables. For the curious reader, we recommend to read Appendix \ref{app: midpoint} where we show how these differences on large scales between the midpoint and the geometric approximations are small when computing the angular correlations.

\section{Impact on weak lensing observables}
\label{sec: Impact on weak lensing observables}

In this section we compute the angular power spectrum in cosmic lensing. First, we calculate the angular correlation function~$w(\theta)$ of the cosmic convergence field, galaxy clustering and galaxy-galaxy lensing, c.f. \cite{Chisari_2019}. As we mentioned above in Sec. \ref{sec1}, they present a general form
\begin{equation}\label{eq:w}
    w^{(i,j)}_{ab}(\theta)
    = \iint_{0}^{\infty}  dx_1 \, dx_2 \, f^i_a(x_1) f^j_b(x_2) \, \xi(r_{12}; t(x_1), t(x_2)) \;, 
\end{equation}
where $(i,j)$ corresponds to the redshift bins, $a$ and $b$ refer to the fields, $x$ is comoving distance, $f^k_c(x)$ are the filters and~$\xi(r_{12}; t(x_1), t(x_2))$ is the unequal-time matter correlation function for the separation
\begin{equation}
    r_{12}
    = \sqrt{x_1^2 + x_2^2 - 2x_1x_2 \cos\theta}
\end{equation}
at cosmic times (or redshift) corresponding to~$x_1$ and~$x_2$ ---c.f. Fig. \ref{fig:Look-back time}. The filters $f^k_c(x)$ for the $k$-th redshift bin depend on the selection function of the corresponding galaxy survey and the field, $c$. The main fields we work with in weak lensing are
\begin{itemize}
    \item \textbf{Cosmic convergence}: given the three-dimensional matter density contrast field~$\delta(\vec{r}; t)$ in comoving coordinates~$\vec{r}$ and cosmic time~$t$, the convergence~$\kappa(\vec{\Theta})$ in direction~$\vec{\Theta}$ is the integral
    \begin{equation}
    \label{eq:kappa}
      \kappa ^i(\vec{\Theta})
      =  \int_{0}^{\infty} \! f^i_\kappa(x) \, \delta(x\vec{\Theta}; t(x)) \, dx
    \end{equation}
    with the filter defined as
    \begin{equation}\label{eq: f_kappa}
        f^i_\kappa(x) = \frac{3H_0^2\Omega_m}{2c^2}\frac{q_i(x) \, x}{a(x)}
    \end{equation}
    where~$H_0$ and~$\Omega_m$ are the cosmological parameters, $c$ is the speed of light, $x$ is comoving distance, $a(x)$ is the scale factor corresponding to~$x$, and $q_i(x)$ is the lensing efficiency given by
    \begin{equation}
    \label{eq:q}
        q_i(x)
        = \int_{x}^{\infty} \! \frac{x' - x}{x'} \, n_i(x') \, dx'
    \end{equation}
    for the distribution~$n_i(x)$ of observed sources within the $i$-th redshift bin.
    \item \textbf{Galaxy number density}: given the three-dimensional matter density contrast field~$\delta(\vec{r}; t)$, the galaxy density field~$\delta_g(\vec{\Theta}; t)$ is the integral
    \begin{equation}
    \label{eq:dg}
      \delta_g^i(\vec{\Theta}; t)
      =  \int_{0}^{\infty} \! f^i_g(x) \, \delta(x\vec{\Theta}; t(x)) \, dx
    \end{equation}
    with the filter defined as
    \begin{equation}\label{eq: f_g}
        f^i_g(x) = \frac{1}{c}b(x) n_i(x) H(x)
    \end{equation}
    where $b(x)$ is the bias parameter and $H(x)$ the Hubble parameter at a redshift corresponding to a comoving distance $x$.
\end{itemize}
In the following we introduce the main observables to be analysed. For simplicity, in this work we ignore intrinsic alignments, redshift-space distortions and lensing magnification.
\begin{itemize}
    \item {\bf Convergence and cosmic shear}. The convergence~$\kappa$ is the cosmic lensing quantity most directly related to the matter field, of which it is the projection along the line of sight.
    The angular correlation function of cosmic convergence reads
    \begin{equation}\label{eq:w-kappa}
      w^{(i,j)}_{\kappa \kappa}(\theta)
      = \iint_{0}^{\infty} dx_1 \, dx_2 \, f^i_\kappa(x_1) f^j_\kappa(x_2) \, \xi(r_{12}; t_1, t_2) \;, 
    \end{equation}
    using the filters \eqref{eq: f_kappa}.

    In practice, it is not cosmic convergence but cosmic shear that is observable. The two-point statistics are related through their respective angular power spectra,~$C_l^{\kappa\kappa}$ and~$C_l^{\gamma\gamma}$, with
    \begin{equation}
     C^{(i,j)}_{\gamma\gamma}(\ell)
     = \frac{(l-1) \, (l+2)}{l \, (l+1)} \,    C^{(i,j)}_{\kappa\kappa}(\ell) \;.
    \end{equation}
    The results we obtain for cosmic convergence are therefore readily applied to cosmic shear. 
    \item {\bf Galaxy clustering}. The angular correlation function quantifies correlations between galaxy number density fields:
    \begin{equation}\label{eq:w-gg}
        w^{(i,j)}_{gg}(\theta)
        = \iint_{0}^{\infty} dx_1 \, dx_2 \, f^i_g(x_1) f^j_g(x_2) \, \xi(r_{12}; t_1, t_2)\;, 
    \end{equation}
    using the filters \eqref{eq: f_g}.
    
    \item {\bf Galaxy-galaxy lensing}. The angular correlation function quantifies the correlation between the shape of background (or source) and foreground (lens) galaxy number density. In the weak lensing regime, the observed galaxy shape is the sum of an intrinsic (unlensed component) and a shear due to gravitational lensing. For simplicity, we only consider the shear component:
    \begin{equation}\label{eq:w-kg}
      w^{(i,j)}_{\kappa g}(\theta)
     = \iint_{0}^{\infty} dx_1 \, dx_2 \, f^i_\kappa(x_1) f^j_g(x_2) \, \xi(r_{12}; t_1, t_2) \;, 
    \end{equation}
    using the filters \eqref{eq: f_kappa} and \eqref{eq: f_g}.

\end{itemize}

\subsection{Computation of the angular power spectrum}
\label{sec:calculations}
There are different approaches to compute the angular power spectra in cosmic lensing \eqref{eq: angular_power_spectrum_AB}:
\begin{enumerate}
    \item Brute-force calculation:
    \begin{equation} \label{eq:exact}
        \begin{split}
             C^{(i,j)}_{ab}(\ell)
         = \frac{2}{\pi} \int_{0}^{\infty} dk k^2 \iint_{0}^{\infty} dx_1 dx_2 \, f^i_a(x_1)f^j_b(x_2)j_{\ell}(k x_1)  j_{\ell}(k x_2)P\left(k; x_1, x_2\right)\; .
        \end{split}
        \end{equation}
    This involves integrals of highly-oscillatory spherical Bessel functions, making the computation time-consuming with potential sources of inaccuracies. This is not due to numerical issues but due to a mathematical behaviour that we cannot control using sophisticated numerical methods.
    \item Approximations to the unequal-time matter power spectrum, which do not reduce the number of integrals. 
    \begin{itemize}
        \item The geometric approximation \eqref{eq:geometric_mean}:
        \begin{equation} \label{eq:geom_int}
        \begin{split}
             C^{(i,j)}_{ab}(\ell)
         \approx \frac{2}{\pi} \int_{0}^{\infty} dk k^2 & \int_{0}^{\infty} dx_1 \, f^i_a(x_1)j_{\ell}(k x_1) \sqrt{P(k; x_1)} \\
         \times & \int_{0}^{\infty} dx_2 f^j_b(x_2) j_{\ell}(k x_2)\sqrt{P(k; x_2)} \; .
        \end{split}
        \end{equation}
        \item The new midpoint approximation \eqref{eq:MRA}: 
        \begin{equation} \label{eq:mid_int}
        \begin{split}
             C^{(i,j)}_{ab}(\ell)
         \approx \frac{2}{\pi} \int_{0}^{\infty} dk k^2 \iint_{0}^{\infty} dx_1 dx_2 \, f^i_a(x_1)f^j_b(x_2)j_{\ell}(k x_1)  j_{\ell}(k x_2)P\left(k; x_m\right)\; .
        \end{split}
        \end{equation}
        where $x_m$ corresponds to the radial distance at the midpoint redshift $z_m = \frac{z_1 + z_2}{2}$.
        \end{itemize}

    \item Assumptions about the filters: the Limber's approximation. 
    \begin{itemize}
        \item In harmonic space, one can use the Dirac-delta version of the Limber's approximation where the spherical Bessel functions are approximated to the amplitude of their first peak
        \begin{equation}\label{eq:Limber delta}
            j_{\ell}(kx) \simeq \sqrt{\pi/(2 \nu)}\delta(\nu - kx)
        \end{equation} 
        with $\nu \equiv \ell + 1/2 $, c.f. \cite{Afshordi:2003xu,Chisari:2018vrw}. Some authors also apply the geometric approximation before using equation \eqref{eq:Limber delta} and call that ``Limber's approximation'', c.f. \cite{Lemos:2017arq}. The curious reader can refer to Appendix \ref{app: Limber_Geometric} to understand why the Limber's and the geometric approximation seem synonyms and why their combination  is completely unnecessary. In addition, note that the Dirac-delta approximation reduces the number of integrals to one, whereas the geometric approximation involves a triple integral.
        
        The Limber's equation reads
        \begin{equation}\label{eq:Limber_delta_int}
            C^{(i,j)}_{ab}(\ell)
         \approx \frac{1}{\nu} \int_{0}^{\infty} \! dk k^2 f^i_a(\nu /k)f^j_b(\nu / k) P(k, \nu / k) \; .
        \end{equation}
    
        \item In angular space (original derivation  \cite{1953ApJ...117..134L}): one can write the angular correlation function~\eqref{eq:w} as the integral over the mean radial distance~$x = (x_1 + x_2)/2$ and the radial separation~$R = x_1 - x_2$
        \begin{equation}\label{eq:w-limber-variables}
            w(\theta)
             = \int_{0}^{\infty} \! \int_{-2x}^{2x} \! f_1(x+R/2) \, f_2(x-R/2) \, \xi(r_{12}; t_1, t_2) \, dR \, dx \;,
        \end{equation}
        where the distance between the points in terms of~$x$ and~$r_{12}$ is now 
        \begin{equation}\label{eq:Limber_r12}
            r_{12} = \sqrt{2 x^2 \, (1-\cos\theta) + R^2 \, (1 + \cos\theta)/2}\;.
        \end{equation}
        Limber \cite{1953ApJ...117..134L} introduced the approximation for the integral~\eqref{eq:w-limber-variables} using the assumption \emph{i)} that the filters and correlation function change slowly and can be approximated by their midpoint values,
        \begin{equation}\label{eq:Limber_filter}
         f_1(x+R/2) \, f_2(x-R/2) \, \xi(r_{12}; t_1, t_2)
        \approx f_1(x) \, f_2(x) \, \xi(r_{12}; t) \;,
        \end{equation}
        \emph{ii)} that the angle separation~$\theta$ between the points is small, so that the distance~$r_{12}$ can be approximated as
        \begin{equation}\label{eq:Limber_sep}
         r_{12}
            \approx \sqrt{x^2\theta^2 + R^2} \;;
        \end{equation}
        \emph{iii)} that the integral over~$R$ can be extended over the entire real line, assuming the spatial correlation function falls off fast enough.
        Limber's approximation for the correlation function~\eqref{eq:w-limber-variables} is thus
        \begin{equation}\label{eq:w-limber}
             w_{\rm L}(\theta)
            = \int_{0}^{\infty} \! f_1(x) \, f_2(x) \, \xi_{\rm L}(x\theta; t) \, x\theta \, dx \;, 
        \end{equation}
        where~$\xi_{\rm L}$ is Limber's matter correlation function, defined as
        \begin{equation}\label{eq:xi_limber}
            \xi_{\rm L}(r; t)
            = \frac{1}{r} \, \int_{-\infty}^{\infty} \! \xi(\sqrt{r^2 + R^2}; t) \, dR \;. 
        \end{equation}

    \end{itemize}

\item In this work, we employ a novel approach by computing the angular correlation function in real space,
\begin{equation}\label{eq:xitow}
    w(\theta)
    = \iint_{0}^{\infty} \! f_1(x_1) \, f_2(x_2) \, \xi(r_{12}; t_1, t_2) \, dx_1 \, dx_2 \;,
\end{equation}
which eliminates the issue of integrating over highly oscillatory functions.
We use the inverse Fourier transform of the unequal-time matter power spectrum \eqref{eq:SPT_UETC_EdS} to obtain the unequal-time correlation function in configuration space,
\begin{equation}\label{eq:ptoxi}
    \xi(r; t_1, t_2)
    = \frac{1}{2\pi^2} \int_{0}^{\infty} \! P(k; t_1, t_2) \, \frac{\sin kr}{kr} \, k^2 \, dk \;,
\end{equation}
which can be evaluated efficiently over a logarithmic range of $r$ values using the FFTLog algorithm \citep{2000MNRAS.312..257H}.
To obtain the angular power spectrum~$C_l$ from the angular correlation function~\eqref{eq:xitow}, we use the general relation between the latter and the former,
\begin{equation}
\label{eq:cltow}
    w(\theta)
    = \sum_{l} \frac{2l + 1}{4\pi} \, C_l \, \LegendreP_l(\cos\theta) \;.
\end{equation}
By evaluating the angular correlation function over a grid of $\theta$ values, and truncating this series at a suitable $l_{\max}$, the modes $C_l$ can be recovered by a least squares fit.

We have implemented all of the above steps in the \texttt{corfu}\footnote{\url{https://github.com/ntessore/corfu}} package for Python \citep{Nicolas_Tessore_2020_4268486}.

\end{enumerate}

\subsection{DES and LSST surveys}

In this section we show the results for cosmic lensing, galaxy clustering and galaxy-galaxy lensing using DES Y1 data \cite{Abbott_2018}, and LSST-like 
Y10 data \cite{thelsstdarkenergysciencecollaboration2018lsst}. We compare the results for a range of approaches  (the Limber's \eqref{eq:w-limber}, the geometric \eqref{eq:geom_int} and the midpoint approximations \eqref{eq:mid_int}) against the exact calculation \eqref{eq:exact}. For this we employ the numerical methods derived in our accompanying paper \cite{Tessore_2020b}.

\subsubsection*{DES Y1 redshift distributions}
We obtain all DES Y1 quantities by running the \verb!des-y1-test! in \verb!CosmoSIS!\footnote{https://bitbucket.org/joezuntz/cosmosis}, without intrinsic alignment, c.f Figure \ref{fig:des-y1}. Specifications:
\begin{itemize}
    \item Cosmology: $$\{\Omega_m = 0.2678, H_0 = 67.5, \Omega_b = 0.0483, n_s = 0.965, A_s = 2.1 \times 10^{-9}, w = -1  \}.$$
    \item Convergence filter \eqref{eq: f_kappa}: we use the source redshift distribution of galaxies for four different redshift bins.
    \item Galaxy filter \eqref{eq: f_g}: we use the lens sample of redshift distribution of galaxies and bias for five different redshift bins. The bias parameter at each redshift:
    $$
        b_{DES}(z) = \{1.45, 1.55, 1.65, 1.8, 2.0 \}\; .
    $$
    \begin{figure}[ht]
        \centering
        \includegraphics[width=0.8\textwidth]{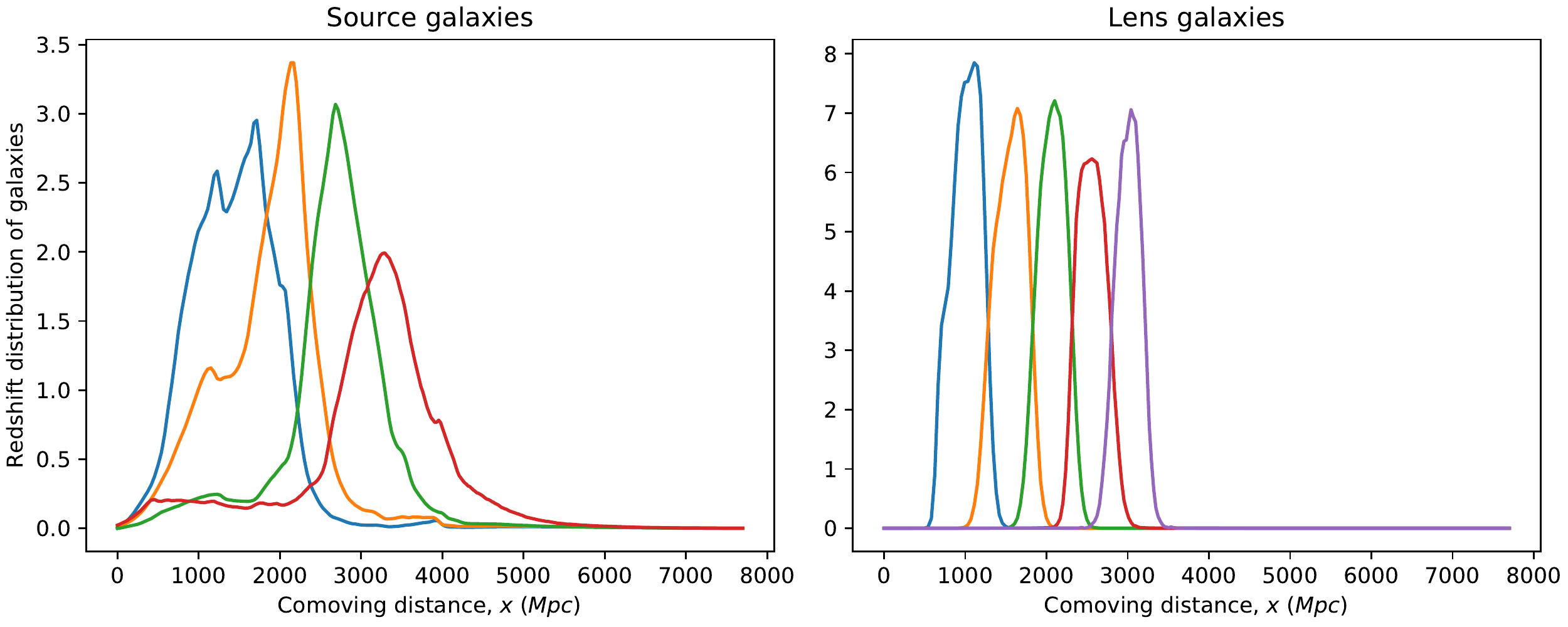}
        \caption{Filters for the source and lens samples for DES-Y1
        redshift distributions.
        } 
        \label{fig:des-y1}
    \end{figure}
\end{itemize}

\subsubsection*{LSST redshift distributions}
We reproduce the redshift distribution of galaxies, 
following the LSST Dark Energy Science Collaboration document \cite{thelsstdarkenergysciencecollaboration2018lsst}
using 10 redshift bins for the Y10 forecast. 
\begin{itemize}
    \item Lens sample. 
    Ten 
    photometric redshift bins $z_{ph} \in (0.2, 1.2)$ with 
    $\Delta z = 0.1$ 
    convolving each bin with a Gaussian photo-z scatter
    \begin{equation}
        \sigma_z = 0.03 (1 + z) \;.
    \end{equation}
    The parametric redshift distribution reads
    \begin{equation}\label{eq:nz}
        n(z) \propto z^2 e^{- (z/z_0)^{\alpha}}  
    \end{equation}
    with parameters 
    $(z_0, \alpha) = (0.28, 0.90)$.
    Then the true distribution of galaxies $n_i(z)$ that fall in the $i$-th photo-z bin is defined
    \begin{equation}
        n_i(z) = \int_{z_{ph}^{i}}^{z_{ph}^{i+1}} d z_{ph} n(z) p(z_{ph}|z)
    \end{equation}
    with a probability distribution $p(z_{ph}|z)$ in $z_{ph}$ at a given redshift, $z$. Instead of solving this integral, we use the error function given in equation (6) by Ma et al. \cite{Ma_2006}.  
    
    Finally, the linear bias parameters are defined:
    \begin{equation}
        b_{LSST}(z_{ph}) = \frac{b}{D(z_{ph})}
    \end{equation}
    with $b = 0.95$.
    $D(z)$ is the normalised linear growth function, which we compute using \verb!SkyPy 0.3! functions \cite{skypy_collaboration_2020_3755531}. 
    \item Source sample. 
    Ten photometric bins $z_{ph}$  defined with equal numbers of source galaxies per bin. 
    This is done using the true redshift distribution, and then the bins are convolved with the photo-z error distribution to make the photo-z distributions: 
    \begin{equation}
        \sigma_z = 0.05 (1 + z) \;.
    \end{equation}
    The parameters in \eqref{eq:nz} are now 
    $(z_0, \alpha) = (0.11, 0.68)$.
\end{itemize}

\subsubsection*{Results}

\begin{figure}[t!]
        \centering
        \includegraphics[width=\textwidth]{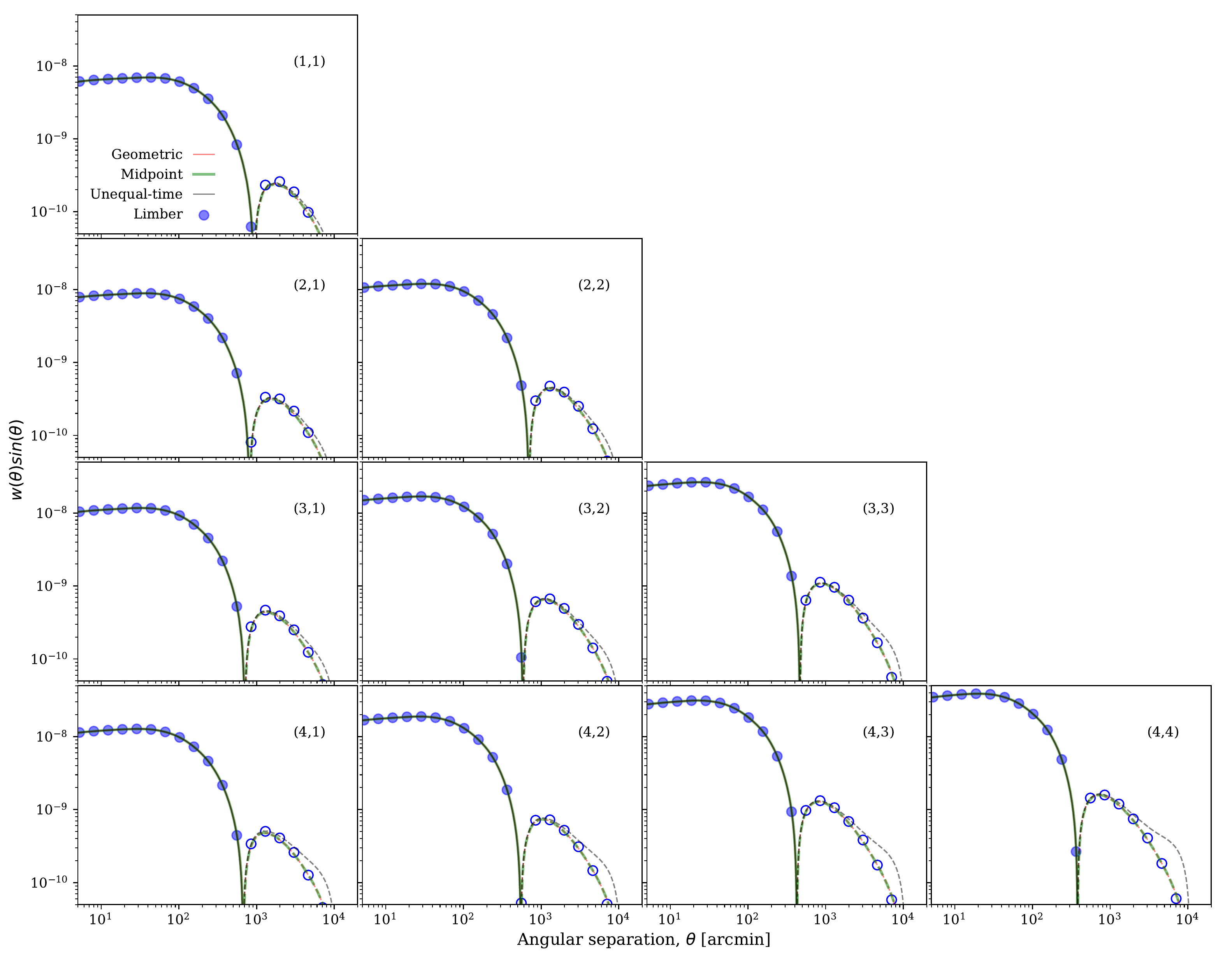}
        \caption{Cosmic lensing. Angular correlation for the whole range of approximations and the fuller calculation at all angular scales. For the unequal-time (black), the geometric (red) and the midpoint (green) approximations, positive values are represented by solid lines and negative values  by dashed lines. The Limber's prediction is represented by blue dots (filled for positive and open for negative values).} 
        \label{fig:lensingw}
    \end{figure}

\begin{figure}[ht]
        \centering
        \includegraphics[width=0.7\textwidth]{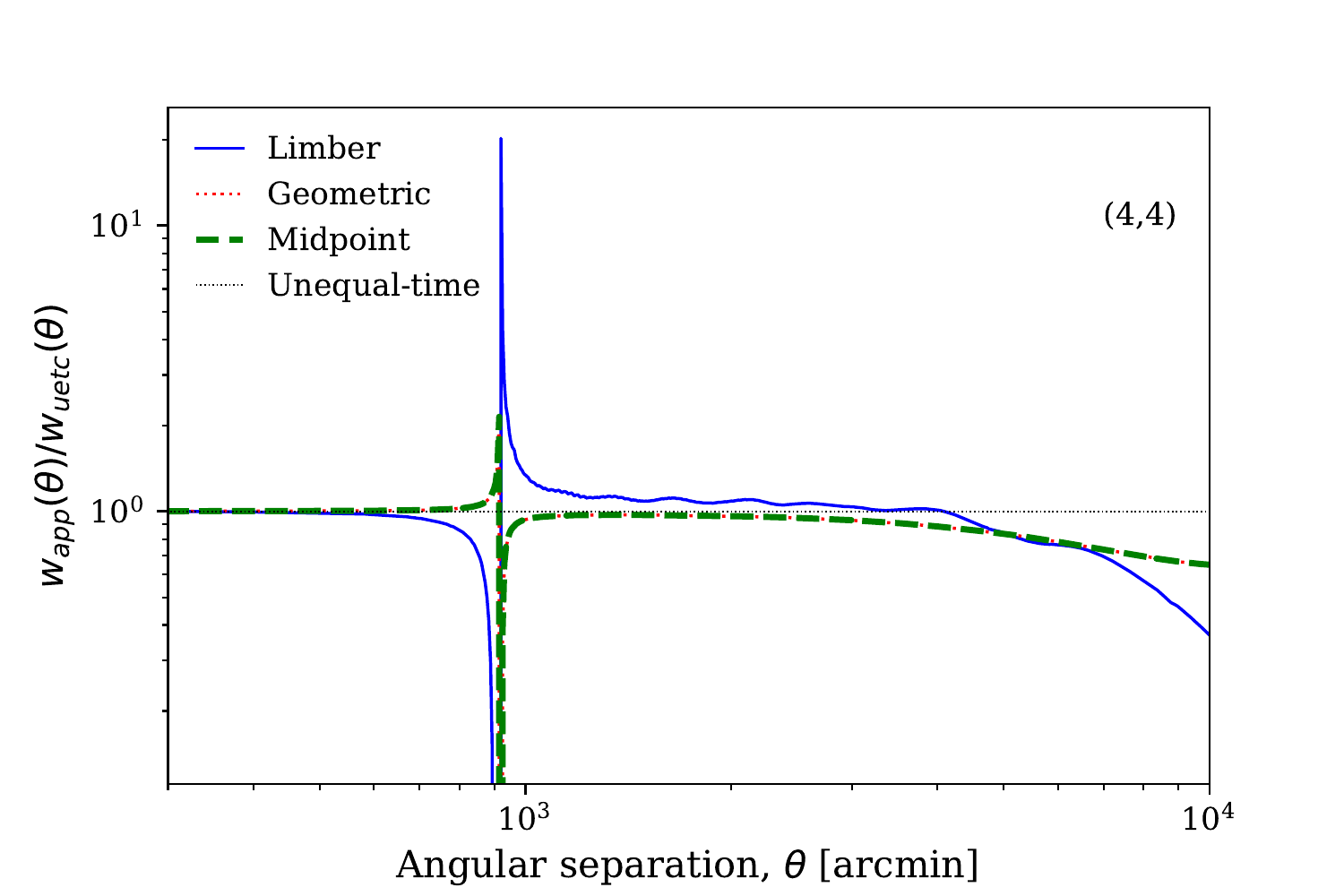}
        \caption{Cosmic lensing. Comparison between the angular correlation function when using an approximation and the fuller calculation at all angular scales for one of the redshift bins, e.g. $(1,1)$. The Limber's approximation does not provide a good prediction on large-angle separation. The midpoint and geometric approximation seems to deal equally well with large-angle correlations, although they deviate from the full unequal-time description. This effect propagates to the angular power spectrum, showing as deviations on small multipoles, $\ell$. The feature at $\theta \approx 9 \times 10^2$ (arcmin) shows the change of sign in the angular correlation function.} 
        \label{fig:lensingw_ratio}
    \end{figure}

\begin{figure}[ht]
        \centering
        \includegraphics[width=\textwidth]{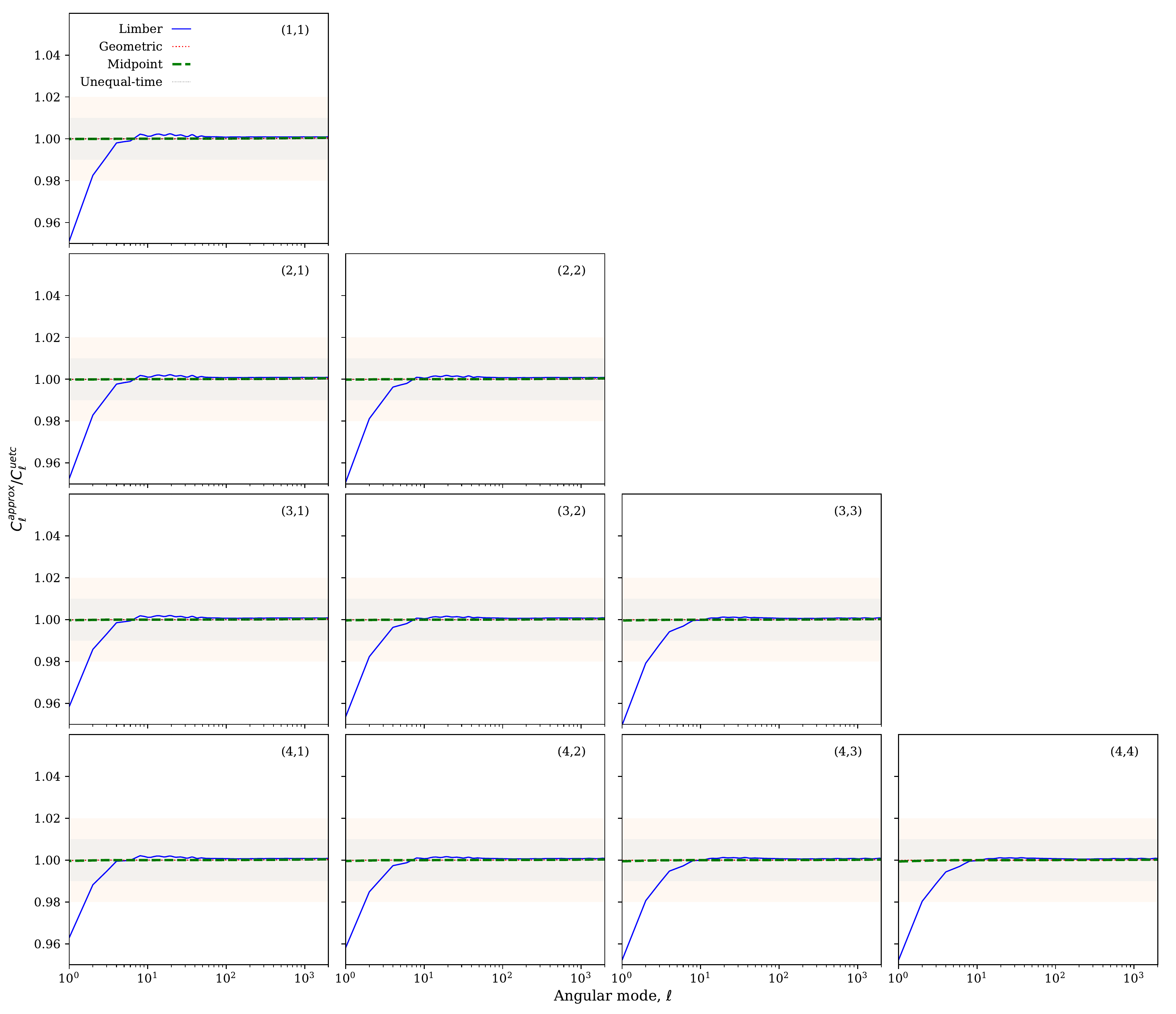}
        \caption{Cosmic lensing. Comparison between the angular power spectrum when using an approximation and the fuller calculation at all angular scales.} 
        \label{fig:lensing}
    \end{figure}

\begin{figure}[ht]
        \centering
        \includegraphics[width=\textwidth]{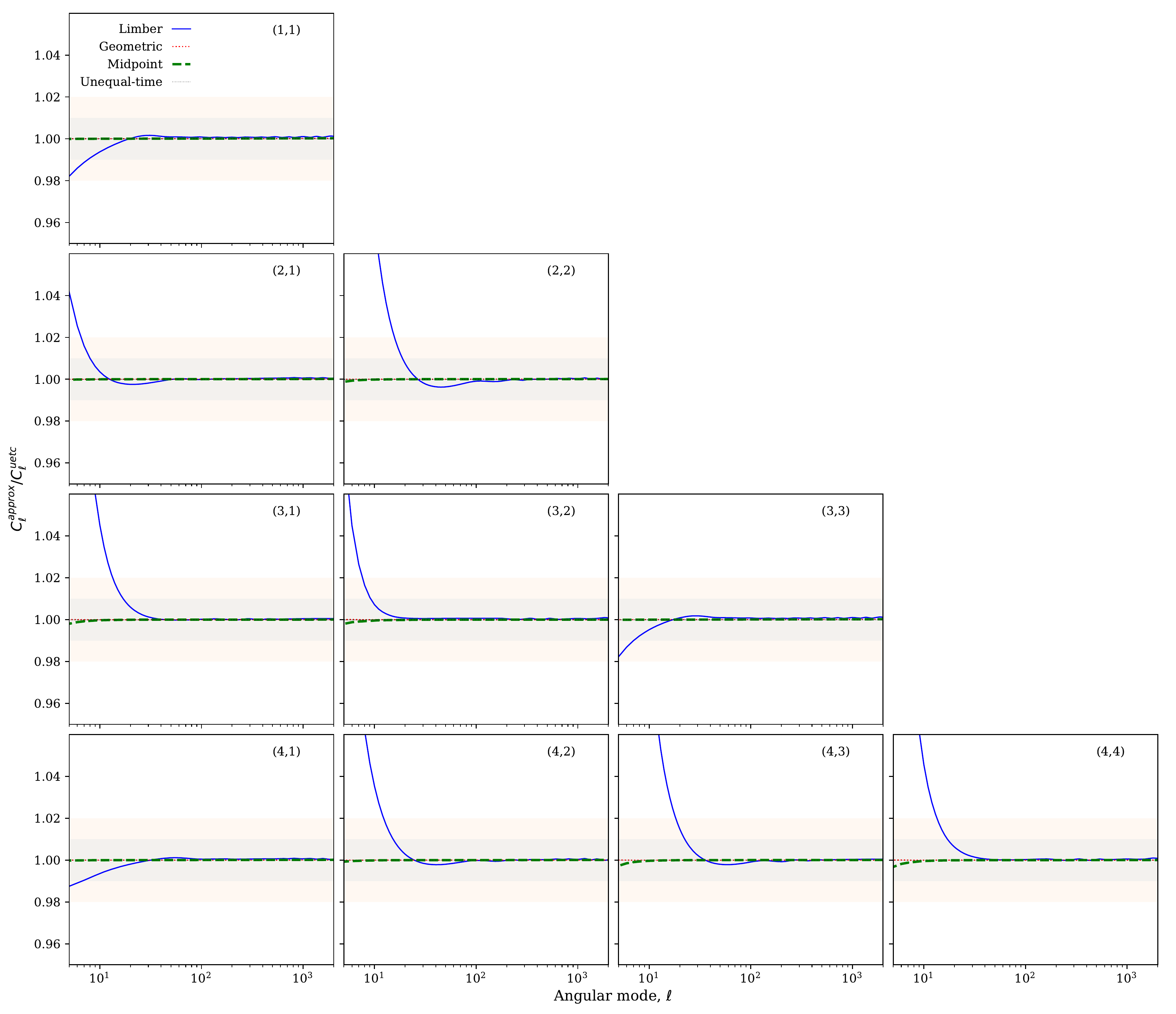}
        \caption{Galaxy-galaxy lensing. Otherwise the same as Figure \ref{fig:des-y1}.} 
        \label{fig:galaxy-galaxy}
    \end{figure}

\begin{figure}[ht]
        \centering
        \includegraphics[width=\textwidth]{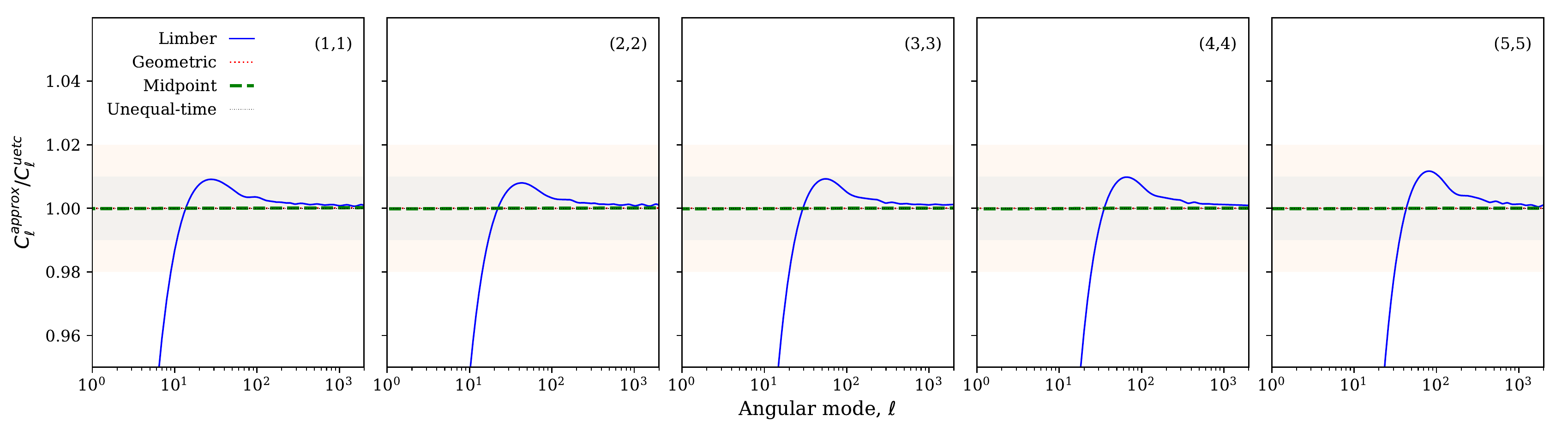}
        \caption{Galaxy clustering. Otherwise the same as Figure \ref{fig:des-y1}.}
        \label{fig:galaxy-clustering}
    \end{figure}

We show the results for cosmic lensing, galaxy-galaxy lensing and galaxy clusters in figures \ref{fig:lensingw}, \ref{fig:lensingw_ratio}, \ref{fig:lensing}, \ref{fig:galaxy-galaxy} and \ref{fig:galaxy-clustering}. We perform the analysis for DES Y1, and LSST-like 
redshift distributions. 
However, to avoid cluttering we only choose to show the results for DES Y1. 
The LSST-like data generates 
similar results to DES Y1, even with narrower redshift bins.  


Simon \cite{Simon:2006gm} and Fang et al. \cite{Fang:2019xat} already warned us that the Limber's approximation could lead to biased predictions for upcoming galaxy surveys like DES Y6,  LSST or Euclid. Here we show that this is true even for DES Y1 data. Likewise, findings by Chisari and Pontzen \cite{Chisari:2019tig} and Kitching and Heavens \cite{Kitching:2016xcl} suggested that non-Limber predictions are sufficient using the equal-time correlators because they are insensitive to the small-scale physics. We agree with this statement, our results draw similar conclusions regardless the unequal-time prescription (one-loop standard perturbation theory, the midpoint or the geometric approximation). For larger-angle separations, c.f. Fig. \ref{fig:lensingw} and \ref{fig:lensingw_ratio}, a full unequal-time description seems to provide a more accurate prediction. Those figures show how the Limber's curve deviates on larger angles and this effect propagates and causes the divergence of the angular power spectrum on small $\ell$. The geometric and midpoint curves also deviates from the unequal-time prediction on large-angle separations.

 Figures  \ref{fig:lensing}, \ref{fig:galaxy-galaxy} and \ref{fig:galaxy-clustering} show the comparison between the angular power spectrum when using one of these approximations (Limber, geometric or midpoint) with respect to the exact calculation at all angular scales using the unequal-time matter power spectrum. The most striking conclusion is the fact that the results from the geometric and midpoint approximations seem to mimic perfectly all the features from the exact calculation. This means that all the features that we observed at the level of the matter power spectrum, Figure \ref{fig: mean approx}, get washed out by the integrals over the line of sight 
(see Appendix \ref{app: midpoint} for details).
Nonetheless, there is a tiny difference for very large-angle separations for galaxy-galaxy lensing and for the farthest redshift bins, for example $(4,4)$. We also observe is that the Limber's approximation deviates from the exact calculation on large angle separations, $\ell < 10$. This is something that we already knew, this effect was predicted by Simon \cite{Simon:2006gm}. This deviation is of order $\sim 2\%$ maximum for cosmic lensing. For galaxy-galaxy lensing the minimum deviation is $\sim 1\%$ and it can be very large for some redshift bins. The biggest effect is found for galaxy clustering.

\section{Conclusions}
\label{sec: Conclusion}

In this paper we have advanced in the field of high precision cosmology, working on the theory side to match the demanding accuracy of upcoming galaxy surveys. These upcoming high-precision observations will have a considerable impact on weak lensing measurements. Therefore, we have devoted our efforts to compute the angular power spectrum exactly at all angular scales using higher-order unequal-time matter power spectrum in perturbation theory. Remember that we define equal-time correlators as the correlation between fields at the same time slice. Likewise, unequal-time correlators measure the correlation between fields at different times slices or redshift.

Until now the most successful approach to compute the angular power spectrum was the Limber's approximation. Its success resides on the reduction of a triple integration to a single integral over the $k$ range. In doing so, it also reduces the complexity at the level of the matter power spectrum as only the equal-time correlator is needed. In Fourier space, the exact calculation involves products of spherical Bessel functions that are highly oscillatory. This makes the computation numerically expensive and highly difficult. Nonetheless, many authors have already shown that such approximation will lead to biased predictions of the cosmological parameters with the upcoming galaxy surveys. Our analysis not only did support this statement, but also concluded that this is true for current data, such as DES Y1.

We computed the one-loop unequal time matter power spectrum using standard perturbation theory and effective field theory to deal with non-linear physics. We have compared with the traditional geometric approximation, concluding that effective field theory breaks at larger scales and tends to give a higher error than the prediction from standard perturbation theory. For these reasons, we continued the rest of our analysis within the standard formalism. We also presented a new unequal-time prescription, the midpoint approximation. We arbitrarily chose our definition of the mean redshift as the average between two different redshift slices. We explained that one can make many other definitions of the mean redshift, with the most natural choice being the mean on comoving distance. However, the same conclusions of this analysis apply regardless of such definition. We also conclude that, at the level of the matter power spectrum, 
the geometric approximation is much better on very large scales, whereas the midpoint approximation seems to equal the unequal-time prediction at some scale that depends on the mean redshift.

In our final section, we showed our results for cosmic lensing, galaxy clustering and galaxy-galaxy lensing. We have used data from DES Y1 and LSST-like 
Y10 for a whole range of approaches (the Limber's approximation, and the geometric and the midpoint approximations at the level of the power spectrum), comparing against the exact calculation. To compute these quantities we have used the numerical methods derived in our accompanying paper \cite{Tessore_2020b} where instead of computing the results in Fourier space, we compute the angular correlations in real space.

In the following, we present a list summarising the main results and conclusions of our work. At the level of the matter power spectrum:
\begin{itemize}
    \item We use the Quijote simulations to find the best value of the counterterms at every redshift available and we parametrise their time evolution. This was never done in the literature before.
    \item We assess the performance of the unequal-time matter power spectrum and we explore the effects coming from non-linear physics and unequal-time correlators. We conclude that the effective field theory framework  does not provide an improved description of the non-linear unequal-time power spectrum and we introduce the new midpoint approximation.
    \item The geometric approximation is better on very large scales and the midpoint approximation show some features that might predict more accurately mild non-linear physics. However, these features do not propagate when computing weak lensing observables. 
\end{itemize}
We showed how neither the effective field theory formalism nor the standard framework provide a completely satisfactory prediction of the matter power spectrum along the line of sight. One reason for this might be the assumption of a homogeneous and isotropic Universe when we define two-point statistics. Observing within our past light-cone breaks homogeneity along the line of sight, and only the spherical symmetry on the two-dimensional sky is preserved. This is worth investigating in the future. One promising alternative could be kinetic field theory \cite{Bartelmann_2019}. This theory allows the straightforward computation of highly-accurate non-linear unequal-time correlators on small scales without free fitting parameters, infinite loop corrections or N-body simulations. We also hope that our work motivates simulation groups to investigate alternative methods to include continuous information along the radial direction, so that we can draw strong conclusions on the best formalism.\\

Finally, the implications on weak lensing are:
\begin{itemize}
    \item The results from the geometric and midpoint approximations mimic perfectly all the features from the exact calculation, washing out any difference at the level of the matter power spectrum.
    \item The particular choice of the midpoint redshift has negligible impact compared to the projection method. In other words, no matter your prescription of the unequal-time matter power spectrum (as long as the correlation functions have unequal-time information), once the integrals are computed over the entire angular range, the angular correlations would be very similar. To compute these integrals, one would need to drop Limber and use our \texttt{corfu} method \citep{Nicolas_Tessore_2020_4268486}.
    \item  We also observe that the result from the Limber's approximation deviates from the exact calculation on large-angle separations, $\ell < 10$.  This deviation is of order $\sim 2\%$ maximum for cosmic lensing. For galaxy-galaxy lensing the minimum deviation is $\sim 1\%$, being very large for some redshift bins. The biggest effect is found for galaxy clustering.
\end{itemize}
Fang et al. \cite{Fang:2019xat}, Chisari and Pontzen \cite{Chisari_2019} and Kitching and Heavens \cite{Kitching:2016xcl} already explained that full-sky observables need to be modelled beyond Limber, e.g. intrinsic galaxy alignments and galaxy shapes. Here we showed that this is true not only for upcoming data, but also for current DES Y1 data. It would be  interesting to see the impact on the prediction of cosmological parameters and check whether it alleviates some of the tensions between galaxy surveys. Although this is out of the scope of our present work, it would be worth investigating in the future.

Our final product is the publicly available python package \verb!unequalpy! \cite{lucia_fonseca_de_la_bella_2020_4268314} with functionality to reproduce all the results and analysis presented in this paper.

\begin{acknowledgments}
We would like to acknowledge all of the insightful comments from our colleagues at the University of Manchester and the SkyPy Collaboration. We specially thank I. Harrison and J. P. Cordero for their insights on Bayesian analysis and redshift distributions. We also thank N. MacCrann for his help with \verb!CosmoSIS!. We also acknowledge A. Heavens, E. Chisari and A. Pontzen, and D. Thomas for their insightful comments and feedback.

The preparation of this manuscript was made possible by a number of software packages: \verb!NumPy!, \verb!SciPy! \cite{Scipy_2020}, \verb!Astropy! \cite{Astropy_2018}, \verb!Matplotlib! \cite{Matplotlib_2007}, \verb!IPython/Jupyter! \cite{Jupyter_2007}, \verb!emcee! \cite{2013PASP..125..306F}. This research made use of \verb!SkyPy!, a Python package for forward modeling  astronomical surveys \cite{skypy_collaboration_2020_3755531}, and \verb!CosmoSIS! \cite{Cosmosis_2015}. This work produced and made use of \verb!corfu! \cite{Nicolas_Tessore_2020_4268486} and \verb!unequalPy! \cite{lucia_fonseca_de_la_bella_2020_4268314}.

The work reported in this paper has been supported by the
European Research Council in the form of a Consolidator Grant with Grant Agreement No. 681431 (LFdlB, NT, SB).

LFdlB would also like to thank all NHS and key workers around the world for their unstoppable work during the global pandemic. LFdlB would also like to add her support to the ``Black Lives Matter''\footnote{https://blacklivesmatter.com} movement and their fight for human rights against racism inside and outside academia.
\end{acknowledgments}

\appendix

\section{Full-time matter power spectrum}
\label{app: Full-time power spectrum} 
De la Bella {\it et al.}~\cite{delaBella:2017qjy} proved the Einstein-De Sitter approximation to lose percent-level accuracy with respect to the full-time computation of the growth functions. The calculations for the equal-time standard perturbation and effective field theory one-loop matter power spectra can be found in that reference. In this appendix, the reader can find the results for their unequal-time counter-parts. 

\begin{itemize}
\item \textbf{Standard Perturbation Theory.}
Within the full-time approach, the    linear and the one-loop contributions now read
\begin{equation}\label{eq:SPT_UETC}
    P_{\mathrm{\tiny SPT}}(k;z_1,z_2) =  P_{11}(k;z_1,z_2) +  P_{22}(k;z_1,z_2) +  2 P_{13}(k;z_1,z_2)
\end{equation}
with
\begin{subequations}
\begin{align}
    \label{eq:P11u}
    P_{11}(k;z_1,z_2) & = D(z_1)D(z_2) P_{11}(k) \\
    \label{eq:P22u}
    \begin{split}
    P_{22}(k;z_1,z_2) & = \DA(z_1) \DA(z_2) P_{AA}(k) + \Big(\DA(z_1) \DB(z_2) + \DA(z_2) \DB(z_1)\Big) P_{AB}(k)\\
                   & + \DB(z_1)\DB(z_2) P_{BB}(k) 
    \end{split}{}
\end{align}
\begin{equation}\label{eq:P13u}
    \begin{split}
    P_{13}(k;z_1,z_2)  = \frac{1}{2}P^\ast(k) \Bigg\{ & \Big[ D(z_1)\Big(\DD(z_2)-\DJ(z_2)\Big) + D(z_2)\Big(\DD(z_1)-\DJ(z_1)\Big)  \Big] P_D(k)\\
                                     + & \Big[ D(z_1)\DE(z_2) + D(z2)\DE(z_1) \Big] P_E(k)\\
                                     + & \Big[ D(z_1)\Big(\DF(z2)+\DJ(z_2)\Big) + D(z_2)\Big(\DF(z_1)+\DJ(z_1)\Big) \Big] P_F(k)\\ 
                                     + & \Big[ D(z_1)\DG(z_2) + D(z_2)\DG(z) \Big] P_G(k)\\ 
                                     + & \frac{1}{2}\Big[ D(z_1)\DJ(z_2) + D(z2)\DJ(z_1) \Big]\Big[ P_{J2}(k) - 2 P_{J1}(k) \Big] \Bigg\}.
    \end{split}
\end{equation}    
\end{subequations}
The quantities $P_i$ appearing in these expressions can be found in de la Bella $et$ $al.$ \citep{delaBella:2017qjy}. In order to compare with the geometric approximation, one would need to square the expression above. 

    \item \textbf{Effective Field Theory.}
Within the full-time approach, the effective field theory unequal-time contribution reads as equation \eqref{eq:EFT_UETC_EdS} but using equation \eqref{eq:SPT_UETC} instead of the Einstein--de Sitter version. The same applies for the expressions for the squared equal-time power spectrum and the squared of the unequal-time counterpart. Note that using the full time dependence of the non-linear growth functions, the best fit values of the counterterms might vary.
\end{itemize}

\section{On why the Limber's and the geometric approximations seem synonyms}
\label{app: Limber_Geometric}
It is common to find in the literature the use of the geometric approximation combined with the Limber's approximation. Some authors decide to apply it before using Limber in its Dirac-delta version \eqref{eq:Limber delta}, c.f \cite{Lemos:2017arq}. However, this is totally unnecessary. When applying the Limber's approximation, the use of the geometric approximation is unnecessary.

Let us write the angular power spectrum of the weak lensing potential, $\phi$, for two different redshift bins, $i$ and $j$:
\begin{equation}
        C^{(i,j)}_{\phi \phi}(\ell)
        = \frac{2}{\pi} \int_{0}^{\infty} \frac{dk}{k^2} \, \iint_{0}^{\infty}  dx_1 \, dx_2 \, f^i_{\phi}(x_1) f^j_{\phi}(x_2)\, j_{\ell}(k x_1) j_{\ell}(k x_2)\, P(k; x_1, x_2) \;. 
\end{equation}
By not using the geometric approximation \eqref{eq:geometric_mean} and simply applying the Limber's approximation in the form \eqref{eq:Limber delta}, the equation above yields
\begin{equation}\label{eq:LimberD}
            C^{(i,j)}_{\phi \phi}(\ell)
         \approx \int_{0}^{\infty} \! \frac{dk}{\nu k^2} f^i_{\phi}(\nu /k)f^j_{\phi}(\nu / k) P(k; \nu/k,\nu/k)
\end{equation}
and from \eqref{eq:geometric_mean_squared}, we are safe to say that
\begin{equation}\label{eq:same_t}
\begin{split}
    P(k; \nu/k,\nu/k) = & \sqrt{P(k; \nu/k)P(k; \nu/k)}\\
                      = & P(k; \nu/k)
\end{split}
\end{equation}
is exact. Therefore, equation \eqref{eq:LimberD} reads
\begin{equation}
            C^{(i,j)}_{\phi \phi}(\ell)
         \approx \int_{0}^{\infty} \! \frac{dk}{\nu k^2} f^i_{\phi}(\nu /k)f^j_{\phi}(\nu / k) P(k; \nu / k)\; .
\end{equation}
Note here, equation \eqref{eq:same_t} is not the geometric approximation \eqref{eq:geometric_mean}. It simply comes from measuring the correlation between fields at the same time slice, and therefore it is exact. It naturally comes out from using the Limber's approximation. Therefore, this result is subjected to all its assumptions and there is no need to impose any extra approximation on the unequal-time correlator.

\section{On why the midpoint and the geometric approximations have similar angular power spectra}
\label{app: midpoint}

\begin{figure}[ht]
        \centering
        \includegraphics[width=0.8\linewidth]{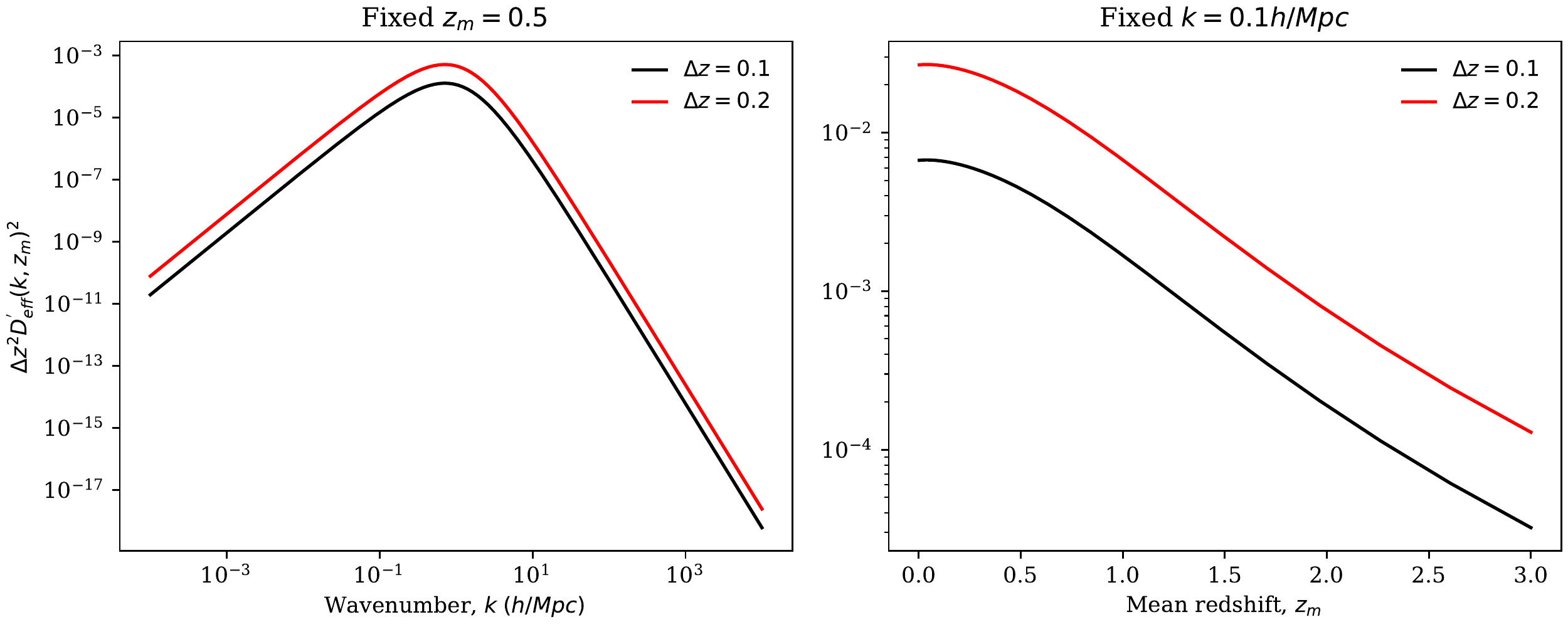}
        \caption{The the difference between the geometric and midpoint approximations depends on the redshift width for a fixed mean redshift, equation \eqref{eq:deltaD}.} 
        \label{fig:deltap} 
\end{figure}

\begin{figure}[ht]
        \centering
        \includegraphics[width=0.6\textwidth]{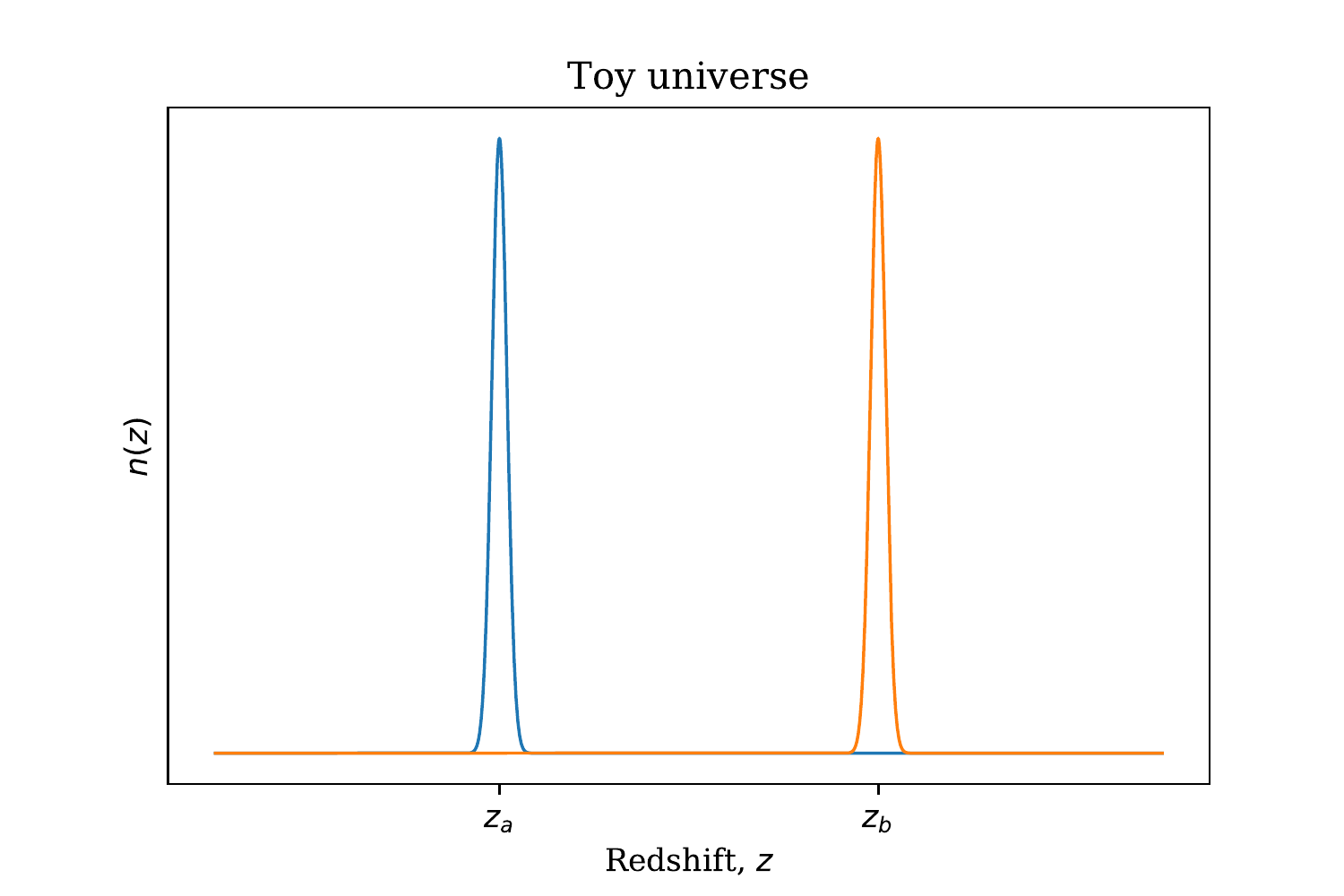}
        \caption{The redshift distribution of galaxies for our toy model.} 
        \label{fig:toy} 
\end{figure}

\begin{figure}[ht]
        \centering
        \includegraphics[width=0.6\textwidth]{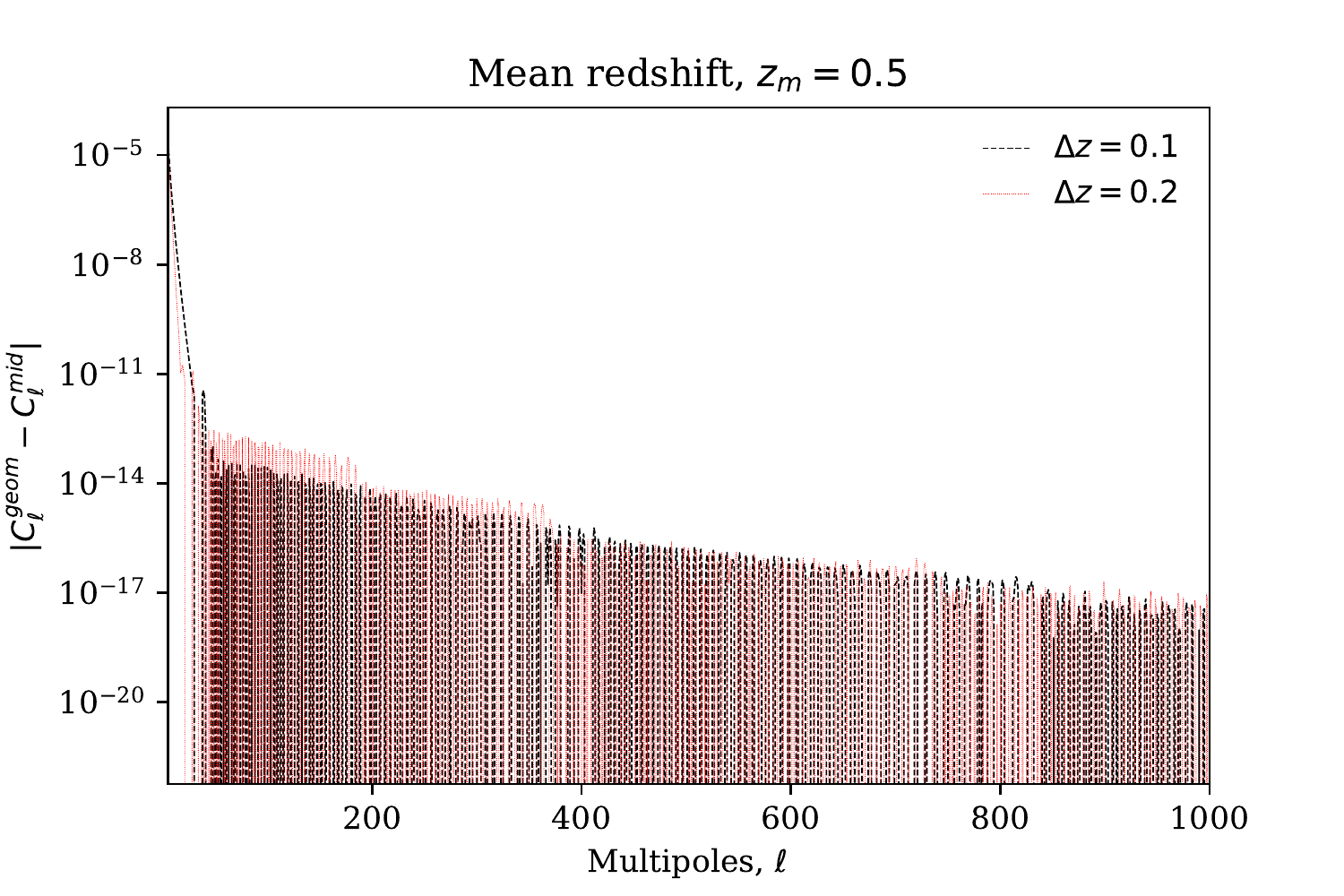}
        \caption{Difference between the angular power spectra using the geometric and the midpoint approximations, equation \eqref{eq:deltaCL}. We see the difference is compatible with zero and oscillations due to the spherical Bessel functions.} 
        \label{fig:deltac}
        \end{figure}

In this section, we show how the features and differences between the geometric and midpoint approximation get washed out after integrating along the line of sight when computing the angular power spectrum. For simplicity, we work with a particular toy model where the distributions of galaxies are given by Dirac deltas. For this we compute the effective growth functions
\begin{itemize}
    \item Equal-time version 
    \begin{equation}\label{eq:Deff_etc}
        D_{eff}(k, z)^2 = \frac{P(k,z)}{P(k, 0)}
    \end{equation} 
    \item Unequal-time version
    \begin{equation}\label{eq:Deff_uetc}
        D_{eff}(k, z_1, z_2)^2 = \frac{P(k,z_1, z_2)}{P(k, 0)} \; .
    \end{equation}
\end{itemize}
Then the difference between the midpoint and geometric approximations at the level of the power spectrum becomes
\begin{equation}\label{eq:deltap}
    |P_{geom} - P_{mid}| = |D_{eff}(k, z_1)D_{eff}(k,z_2) - D_{eff}^2(k, z_m)| P(k, 0) \;.
\end{equation}
We now perform a Taylor expansion of the effective growth functions around the mean redshift value, $z_m$:
\begin{equation}
    D_{eff}(k,z) \approx D_{eff}(k,z_m) + (z - z_m) \frac{\partial D_{eff}(k,z)}{\partial z} \bigg|_{z_m} + (z - z_m)^2 \frac{\partial^2 D_{eff}(k,z)}{\partial z^2} \bigg|_{z_m}
\end{equation}
Therefore, at leading order
\begin{equation}\label{eq:deltaD}
    |D_{eff}(k, z_1)D_{eff}(k,z_2) - D_{eff}^2(k, z_m)| \approx \left( \Delta z D_{eff}'(k,z_m) \right)^2
\end{equation}
with $'\equiv \partial / \partial z$. Remember, $z_m = (z_1 + z_2) /2$ and $\Delta z = (z_1 - z_2) /2$. Figure \ref{fig:deltap} shows that the difference between these approximations \eqref{eq:deltap} only depends on the redshift width for a fixed mean redshift. This is reasonable, given that the midpoint approximation does not use information from the redshift separations. 
If we now consider a toy model, c.f. \ref{fig:toy} where the redshift distribution of galaxies follow this law
\begin{equation}
    n(z) = \delta _D (z - z_a) + \delta _D (z - z_b)
\end{equation}
in a way that $z_m = (z_a + z_b) /2$ and $\Delta z = (z_a - z_b) /2$.
When we compute the angular power spectra for both approximations, the difference reads
\begin{equation}\label{eq:deltaCL}
    |C_{geom}(\ell) - C_{mid}(\ell)| \approx \left( \Delta z D_{eff}'(k,z_m) \right)^2 \int \frac{d k}{k^2} j_{\ell}(k z_a) j_{\ell}(k z_b) P(k, 0) 
\end{equation}
which is a very small quantity, since both factors are small themselves. This is shown in Figure \ref{fig:deltac}, where the oscillations come from the product of the spherical Bessel functions. We can conclude that any big differences between approximations at the level of the matter power spectrum, seems to have little impact when integrating along the line of sight.

\clearpage
\phantomsection
\renewcommand{\bibname}{References}
\bibliographystyle{JHEP}
\bibliography{main}

\providecommand{\href}[2]{#2}\begingroup\raggedright\begin{thebibliography}{10}

\bibitem{Kitching:2016xcl}
T.~D. Kitching and A.~F. Heavens, \emph{{Unequal-Time Correlators for
  Cosmology}}, \href{https://doi.org/10.1103/PhysRevD.95.063522}{\emph{Phys.
  Rev.} {\bfseries D95} (2017) 063522}
  [\href{https://arxiv.org/abs/1612.00770}{{\ttfamily 1612.00770}}].

\bibitem{Tessore_2020b}
N.~Tessore and L.~F. de~la Bella, \emph{{Cosmic lensing two-point functions
  with exact projection}}, .

\bibitem{Chisari:2019tig}
N.~E. Chisari and A.~Pontzen, \emph{{Unequal time correlators and the
  Zel’dovich approximation}},
  \href{https://doi.org/10.1103/PhysRevD.100.023543}{\emph{Phys. Rev.}
  {\bfseries D100} (2019) 023543}
  [\href{https://arxiv.org/abs/1905.02078}{{\ttfamily 1905.02078}}].

\bibitem{1953ApJ...117..134L}
D.~N. {Limber}, \emph{{The Analysis of Counts of the Extragalactic Nebulae in
  Terms of a Fluctuating Density Field.}},
  \href{https://doi.org/10.1086/145672}{\emph{APJ} {\bfseries 117} (1953) 134}.

\bibitem{1954ApJ...119..655L}
D.~N. {Limber}, \emph{{The Analysis of Counts of the Extragalactic Nebulae in
  Terms of a Fluctuating Density Field. II.}},
  \href{https://doi.org/10.1086/145870}{\emph{"APJ"} {\bfseries 119} (1954)
  655}.

\bibitem{1973ApJ...185..413P}
P.~J.~E. {Peebles}, \emph{{Statistical Analysis of Catalogs of Extragalactic
  Objects. I. Theory}}, \href{https://doi.org/10.1086/152431}{\emph{"APJ"}
  {\bfseries 185} (1973) 413}.

\bibitem{Kaiser_1998}
N.~Kaiser, \emph{Weak lensing and cosmology},
  \href{https://doi.org/10.1086/305515}{\emph{The Astrophysical Journal}
  {\bfseries 498} (1998) 26–42}.

\bibitem{Lemos:2017arq}
P.~Lemos, A.~Challinor and G.~Efstathiou, \emph{{The effect of Limber and
  flat-sky approximations on galaxy weak lensing}},
  \href{https://doi.org/10.1088/1475-7516/2017/05/014}{\emph{JCAP} {\bfseries
  05} (2017) 014} [\href{https://arxiv.org/abs/1704.01054}{{\ttfamily
  1704.01054}}].

\bibitem{Simon:2006gm}
P.~Simon, \emph{{How accurate is Limber's equation?}},
  \href{https://doi.org/10.1051/0004-6361:20066352}{\emph{Astron. Astrophys.}
  {\bfseries 473} (2007) 711}
  [\href{https://arxiv.org/abs/astro-ph/0609165}{{\ttfamily
  astro-ph/0609165}}].

\bibitem{Kitching_2017}
T.~D. Kitching, J.~Alsing, A.~F. Heavens, R.~Jimenez, J.~D. McEwen and
  L.~Verde, \emph{The limits of cosmic shear},
  \href{https://doi.org/10.1093/mnras/stx1039}{\emph{Monthly Notices of the
  Royal Astronomical Society} {\bfseries 469} (2017) 2737–2749}.

\bibitem{LoVerde:2008re}
M.~LoVerde and N.~Afshordi, \emph{{Extended Limber Approximation}},
  \href{https://doi.org/10.1103/PhysRevD.78.123506}{\emph{Phys. Rev. D}
  {\bfseries 78} (2008) 123506}
  [\href{https://arxiv.org/abs/0809.5112}{{\ttfamily 0809.5112}}].

\bibitem{Fang:2019xat}
X.~Fang, E.~Krause, T.~Eifler and N.~MacCrann, \emph{{Beyond Limber: Efficient
  computation of angular power spectra for galaxy clustering and weak
  lensing}}, \href{https://doi.org/10.1088/1475-7516/2020/05/010}{\emph{JCAP}
  {\bfseries 05} (2020) 010}
  [\href{https://arxiv.org/abs/1911.11947}{{\ttfamily 1911.11947}}].

\bibitem{lucia_fonseca_de_la_bella_2020_4268314}
L.~F. de~la Bella, \emph{unequalpy},  Nov., 2020.
\newblock 10.5281/zenodo.4268314.

\bibitem{Carrasco:2012cv}
J.~J.~M. Carrasco, M.~P. Hertzberg and L.~Senatore, \emph{{The Effective Field
  Theory of Cosmological Large Scale Structures}},
  \href{https://doi.org/10.1007/JHEP09(2012)082}{\emph{JHEP} {\bfseries 09}
  (2012) 082} [\href{https://arxiv.org/abs/1206.2926}{{\ttfamily 1206.2926}}].

\bibitem{delaBella:2017qjy}
L.~F. de~la Bella, D.~Regan, D.~Seery and S.~Hotchkiss, \emph{{The matter power
  spectrum in redshift space using effective field theory}},
  \href{https://doi.org/10.1088/1475-7516/2017/11/039}{\emph{JCAP} {\bfseries
  1711} (2017) 039} [\href{https://arxiv.org/abs/1704.05309}{{\ttfamily
  1704.05309}}].

\bibitem{Bernardeau:2001qr}
F.~Bernardeau, S.~Colombi, E.~Gaztanaga and R.~Scoccimarro, \emph{{Large scale
  structure of the universe and cosmological perturbation theory}},
  \href{https://doi.org/10.1016/S0370-1573(02)00135-7}{\emph{Phys. Rept.}
  {\bfseries 367} (2002) 1}
  [\href{https://arxiv.org/abs/astro-ph/0112551}{{\ttfamily
  astro-ph/0112551}}].

\bibitem{skypy_collaboration_2020_3755531}
S.~Collaboration, \emph{Skypy},  Apr., 2020.
\newblock 10.5281/zenodo.4071945.

\bibitem{Villaescusa-Navarro:2019bje}
F.~Villaescusa-Navarro et~al., \emph{{The Quijote simulations}},
  \href{https://arxiv.org/abs/1909.05273}{{\ttfamily 1909.05273}}.

\bibitem{Chisari_2019}
N.~E. Chisari, D.~Alonso, E.~Krause, C.~D. Leonard, P.~Bull, J.~Neveu et~al.,
  \emph{Core cosmology library: Precision cosmological predictions for lsst},
  \href{https://doi.org/10.3847/1538-4365/ab1658}{\emph{The Astrophysical
  Journal Supplement Series} {\bfseries 242} (2019) 2}.

\bibitem{Afshordi:2003xu}
N.~Afshordi, Y.-S. Loh and M.~A. Strauss, \emph{{Cross - correlation of the
  Cosmic Microwave Background with the 2MASS galaxy survey: Signatures of dark
  energy, hot gas, and point sources}},
  \href{https://doi.org/10.1103/PhysRevD.69.083524}{\emph{Phys. Rev. D}
  {\bfseries 69} (2004) 083524}
  [\href{https://arxiv.org/abs/astro-ph/0308260}{{\ttfamily
  astro-ph/0308260}}].

\bibitem{Chisari:2018vrw}
{\scshape LSST Dark Energy Science} collaboration, \emph{{Core Cosmology
  Library: Precision Cosmological Predictions for LSST}},
  \href{https://doi.org/10.3847/1538-4365/ab1658}{\emph{Astrophys. J. Suppl.}
  {\bfseries 242} (2019) 2} [\href{https://arxiv.org/abs/1812.05995}{{\ttfamily
  1812.05995}}].

\bibitem{2000MNRAS.312..257H}
A.~J.~S. {Hamilton}, \emph{{Uncorrelated modes of the non-linear power
  spectrum}},
  \href{https://doi.org/10.1046/j.1365-8711.2000.03071.x}{\emph{"MNRAS"}
  {\bfseries 312} (2000) 257}
  [\href{https://arxiv.org/abs/astro-ph/9905191}{{\ttfamily
  astro-ph/9905191}}].

\bibitem{Nicolas_Tessore_2020_4268486}
N.~Tessore and L.~F. de~la Bella, \emph{corfu},  Nov., 2020.
\newblock 10.5281/zenodo.4268486.

\bibitem{Abbott_2018}
T.~Abbott, F.~B. Abdalla, A.~Alarcon, J.~Aleksić, S.~Allam, S.~Allen et~al.,
  \emph{Dark energy survey year 1 results: Cosmological constraints from galaxy
  clustering and weak lensing},
  \href{https://doi.org/10.1103/physrevd.98.043526}{\emph{Physical Review D}
  {\bfseries 98} (2018) }.

\bibitem{thelsstdarkenergysciencecollaboration2018lsst}
T.~L. D. E.~S. Collaboration, R.~Mandelbaum, T.~Eifler, R.~Hložek, T.~Collett,
  E.~Gawiser et~al., \emph{The lsst dark energy science collaboration (desc)
  science requirements document},  2018.

\bibitem{Ma_2006}
Z.~Ma, W.~Hu and D.~Huterer, \emph{Effects of photometric redshift
  uncertainties on weak‐lensing tomography},
  \href{https://doi.org/10.1086/497068}{\emph{The Astrophysical Journal}
  {\bfseries 636} (2006) 21–29}.

\bibitem{Bartelmann_2019}
M.~Bartelmann, E.~Kozlikin, R.~Lilow, C.~Littek, F.~Fabis, I.~Kostyuk et~al.,
  \emph{Cosmic structure formation with kinetic field theory},
  \href{https://doi.org/10.1002/andp.201800446}{\emph{Annalen der Physik}
  {\bfseries 531} (2019) 1800446}.

\bibitem{Scipy_2020}
P.~{Virtanen}, R.~{Gommers}, T.~E. {Oliphant}, M.~{Haberland}, T.~{Reddy},
  D.~{Cournapeau} et~al., \emph{{SciPy 1.0: fundamental algorithms for
  scientific computing in Python}},
  \href{https://doi.org/10.1038/s41592-019-0686-2}{\emph{Nature Methods}
  {\bfseries 17} (2020) 261}
  [\href{https://arxiv.org/abs/1907.10121}{{\ttfamily 1907.10121}}].

\bibitem{Astropy_2018}
A.~M. Price-Whelan, B.~M. Sipőcz, H.~M. Günther, P.~L. Lim, S.~M. Crawford,
  S.~Conseil et~al., \emph{The astropy project: Building an open-science
  project and status of the v2.0 core package},
  \href{https://doi.org/10.3847/1538-3881/aabc4f}{\emph{The Astronomical
  Journal} {\bfseries 156} (2018) 123}.

\bibitem{Matplotlib_2007}
J.~D. {Hunter}, \emph{{Matplotlib: A 2D Graphics Environment}},
  \href{https://doi.org/10.1109/MCSE.2007.55}{\emph{Computing in Science and
  Engineering} {\bfseries 9} (2007) 90}.

\bibitem{Jupyter_2007}
F.~{Perez} and B.~E. {Granger}, \emph{{IPython: A System for Interactive
  Scientific Computing}},
  \href{https://doi.org/10.1109/MCSE.2007.53}{\emph{Computing in Science and
  Engineering} {\bfseries 9} (2007) 21}.

\bibitem{2013PASP..125..306F}
D.~{Foreman-Mackey}, D.~W. {Hogg}, D.~{Lang} and J.~{Goodman}, \emph{{emcee:
  The MCMC Hammer}}, \href{https://doi.org/10.1086/670067}{\emph{{PASP}}
  {\bfseries 125} (2013) 306}
  [\href{https://arxiv.org/abs/1202.3665}{{\ttfamily 1202.3665}}].

\bibitem{Cosmosis_2015}
J.~Zuntz, M.~Paterno, E.~Jennings, D.~Rudd, A.~Manzotti, S.~Dodelson et~al.,
  \emph{Cosmosis: Modular cosmological parameter estimation},
  \href{https://doi.org/10.1016/j.ascom.2015.05.005}{\emph{Astronomy and
  Computing} {\bfseries 12} (2015) 45–59}.

\end{thebibliography}\endgroup

\end{document}